\documentclass[letterpaper]{article}
\usepackage[pass]{geometry}
\usepackage{authblk}
\usepackage{times}
\usepackage[authoryear]{natbib}
\usepackage{url}
\usepackage{graphicx,amsfonts,amsmath}
\usepackage{xcolor}
\usepackage{xargs}

\usepackage{tikz}
\usepackage{pgfplots}

\pgfplotsset{compat=1.5}\graphicspath{{/home/bazzan/pesquisa/teledramaturgia/friends/Plots/}}

\newcommand{\fri}{\textit{Friends}}
\newcommand{\netth}{network theory}
\newcommand{\cd}{community detection}
\newcommand{\nmi}{NMI}

\title{I will be there for you:\\ six friends in a clique\thanks{This is a pre-print of an article submitted to Computational and Applied Mathematics (COAM). 
}
}

\author{Ana L. C. Bazzan}
\affil{Computer Science Institute, UFRGS\\
91501-970 -- P. Alegre, Brazil\\ 
bazzan@inf.ufrgs.br}

\date{}

\begin{document}

\maketitle

\begin{abstract}
Network science has proved useful in analyzing structure and dynamics of social networks in several areas.
This paper aims at analyzing the relationships of characters in \textit{Friends}, a famous sitcom.
In particular, two important aspects are investigated.
First, how are the structure of the communities (groups)? How different methods for community detection perform?
Second, not only static structure of the graphs and causality relationships are investigated, but also temporal aspects.
After all, this show was aired for ten years and thus plots, roles, and friendship patterns among the characters seem to have changed.
Also, this sitcom is frequently associated with distinguishing facts such as: all six characters are equally prominent; it has no dominant storyline; and friendship as surrogate family.
This paper uses tools from network theory to check whether these and other facts can be quantified and proved correct, especially considering the temporal aspect, i.e., what happens in the sitcom along time.
The main findings regarding the centrality and temporal aspects are: patterns in graphs representing different time slices of the show change; overall, degrees of the six friends are indeed nearly the same; however, in different situations (thus graphs), the magnitudes of degree centrality do change; betweenness centrality differs significantly for each character thus some characters are better connectors than others; 
there is a high difference regarding degrees of the six friends versus the rest of the characters, which points to a centralized network; there are strong indications that the six friends are part of  a surrogate family.
As for the presence of groups within the network, methods of different natures were investigated aiming at detecting groups (communities) in networks representing different time slices as well as the network of all episodes. Such methods were compared (pairwise and also using various metrics, including plausibility). The multilevel method performs reasonably in general. Also, it stands out that those methods do not agree very much, resulting in groups that are very different from method to method. 
\end{abstract}

\section*{Introduction}

With the increasing penetration of streaming technology, TV shows and series are becoming more and more popular.
What makes some shows so appealing?
Besides obvious items (plot, cast, cinematography, etc.), the structure of the  network of characters underlying the plot (the social network of the show's plot) can offer some hints too.

The use of \netth\ 
-- which  studies complex interacting systems  represented as graphs -- is a nice way to shed light on questions related to the social network  underlying a TV show.
For example, \cite{Beveridge&Shan2016} investigated who is/are the most central characters in \textit{Game of Thrones}.
This  popular show was also the target of \cite{Jasonov2017} who computed the importance of characters and used them as features or input to a machine learning algorithm in order to predict how likely to die  some characters are.
\cite{Tan+2014} analyzed the character networks of  Stargate and Star Trek and found that their structures are quite similar.

These  studies investigate particular issues related to these shows.
However, 
in none of them the temporal  (not only causal) aspects of the shows were deeply explored. 
Also, in many cases,  the data employed to construct the networks were neither based on the entirety of the episodes, nor manually collected, which means that some parsing or other automated strategy had to be used.
In the case of \cite{Beveridge&Shan2016}, they constructed a graph by including an edge between any two characters whose name appeared within 15 words in the text of the third book (A Storm of Swords); \cite{Jasonov2017} collected  data about available scenes in dialogue (subtitles) format on a fan website (\url{genious.com}) and assumed that  within a scene everyone was then connected with everyone.

In the present study, a broader range of issues (e.g., related to temporal patterns and community structure) of the situation comedy (sitcom) \fri\ is analyzed, spanning through ten seasons.

\fri\ is an American television sitcom created by  David Crane and Marta Kauffman, which was aired on NBC from 1994 to 2004.
\fri\ featured six main characters -- Rachel Green (Jennifer Aniston), Monica Geller (Courteney Cox), Phoebe Buffay (Lisa Kudrow), Joey Tribbiani (Matt LeBlanc), Chandler Bing (Matthew Perry), and Ross Geller (David Schwimmer).
The story unfolds at three main settings: a Manhattan coffeehouse (Central Perk) and the apartments of Monica and Rachel and Joey and Chandler across the hall.

According to \cite{Sternbergh2016}, with the arrival of \fri\ to Netflix, the show is reaching a whole new generation of 20--30 year olds, and its popularity is on the rise.

For the present study, data of each  episode of \fri\ was manually collected based on the actual interactions of  characters in each scene.
An interaction happens when two characters talk (even if  one talks and the other just listens) or touch or have eye contact.
This means that, since not necessarily every character  does interact with all others in a scene, each scene is not a complete graph connecting all characters in it.
Thus, there are some differences between the way  graphs are constructed in \cite{Beveridge&Shan2016}, \cite{Jasonov2017}, and in the present work.
Data was  collected by watching each of the 236 episodes\footnote{Available at \url{https://github.com/anabazzan/friends}}. 
Pairwise interactions were stored in text files that were then  processed using \textit{igraph} \citep{Csardi&Nepusz2006} for \textit{python}\footnote{Available at \url{http://igraph.org/python/}}.
One can either look at graphs for each episode, for all episodes together, or for any particular merge of episodes/situations (e.g., Thanksgiving episodes, all first episodes in each season). 

The main aim is to check whether well known facts about \fri\ -- e.g., that all six characters are equally prominent -- can be quantified and proved correct.
Moreover, what can be said about such facts regarding different contexts or the passing of time? After all, \fri\ aired for ten years and things might have changed.
Finally, how do known methods for community detection perform in the case of this dataset?
Are there similarities to other human social networks?

\section*{\fri\ as inspiration for academic studies}
It is only natural that the sitcom \fri\ has attracted the attention of researchers in the area of Arts and Communication, especially in the late 1990s and the 2000s, when the show was still being aired or had just ended.
However, \fri\ has also been the subject of a myriad of interesting studies, ranging from Social Sciences and Linguistics to Math and Computer Science.
Moreover, the list includes recent work as well, showing that  \fri\ is still  popular.
Some of these 
are: 
L. \cite{Marshall2007}'s thesis  examined representations of friendship, gender, race, and social class in \fri.
P. \cite{Quaglio2009} compared the language of \fri\ to natural conversation, in particular comparing high-frequency linguistic features that characterize conversation to the language of \fri.
T. \cite{Heyd2010} studied the construction ``\textit{you guys}'' as an emerging quasi pronoun for second-person plural address based on dialogue transcriptions of  \fri.
C.-J. \cite{Nan+2015} used a 
deep learning model for face recognition in \fri's videos in order to distinguish the six main characters and establish the social network between them.
\cite{Edwards+2018} compared different extraction methods for social networks in narratives
providing evidence that automated methods of data extraction are reliable for many (though not all)  analyses.

\section*{What  \fri\ is known for}

\fri\ is frequently associated with these facts (especially the first two):
\begin{itemize}
 \item All six characters are equally prominent;
 \item \fri\ is a multistory sitcom  with no dominant storyline; 
 \item Monica likes to consider herself as hostess / mother hen; 
 \item Friendship as surrogate family;
 \item Ross and Rachel have an intermittent relationship;
 \item Chandler and Phoebe had originally been written as more secondary characters, to provide humor when needed;
 \item The writers originally planned a big love story between Joey and Monica.
\end{itemize}

In this article, these  facts are investigate using tools of \netth.
If the six characters in this sitcom are indeed equally prominent, one expects that quantitative measures of their importance in the story confirm this.
If there is no dominant storyline, there should be no prominent character(s) in the episodes (apart from obvious exceptions).
These (and other) characteristics of the  social network of \fri\ are analyzed here, in a trans-disciplinary effort to establish connections between Humanities and Mathematics.

The main characteristics -- e.g.,  multistory with no prominent character -- as well as the reasons behind the popularity of \fri\ have been the subject of various studies,  stemming both from academic circles as well as from daily newspapers, blogs, etc.
Here are some quotations that corroborate the just mentioned facts about this sitcom:

\begin{enumerate}
 \item ``This series has six major characters, three men and three women, who are generally given equal weight across the series. '': K. \cite{Thompson2003},  page 56;
\item 
``Beyond its glamour, "Friends" is widely lauded as the first true "ensemble" show -- a series with no clear star or center, a cast of equals with no authority figure in sight.'': C. \cite{McCarroll2004};
 \item  ``Friends would be a pure ensemble, with none of the six more prominent than any other. "No one had done a true ensemble," [Series creator David] Crane said. 
 The creators felt that utilizing six equal players, rather than emphasizing one or two, would allow for myriad story lines and give the show legs, according to Crane.'': T. \cite{Jischa2004};
\item ``The concept for Friends never deviated from the vision of its creators. 
... a comedy involving young people in a big city coming together to share living expenses. 
This meant they also would share signal events in a memorable period of their lives, not with parents and siblings, but with new, surrogate family members.'': T. \cite{Jischa2004}
\item  ``... when we first pitched the show, although we always said it was an ensemble show, we kind of thought of Phoebe and  Chandler  as a little more secondary; they would provide humor when we need... they gave us much much more than that; they became so central to the ensemble...'':  M. \cite{Kauffmann2004};
\item 
``The writers originally planned a big love story between Joey and Monica [...] The idea of a romantic interest between Ross and Rachel emerged during the period when Kauffman and Crane wrote the pilot script.'': M. \cite{Lauer2004}.
\end{enumerate}

\section*{What can \netth\  say about these facts?}

\begin{figure}[h!]
 \centering
 \includegraphics[trim={0 0 0 2cm},clip,width=0.7\linewidth]{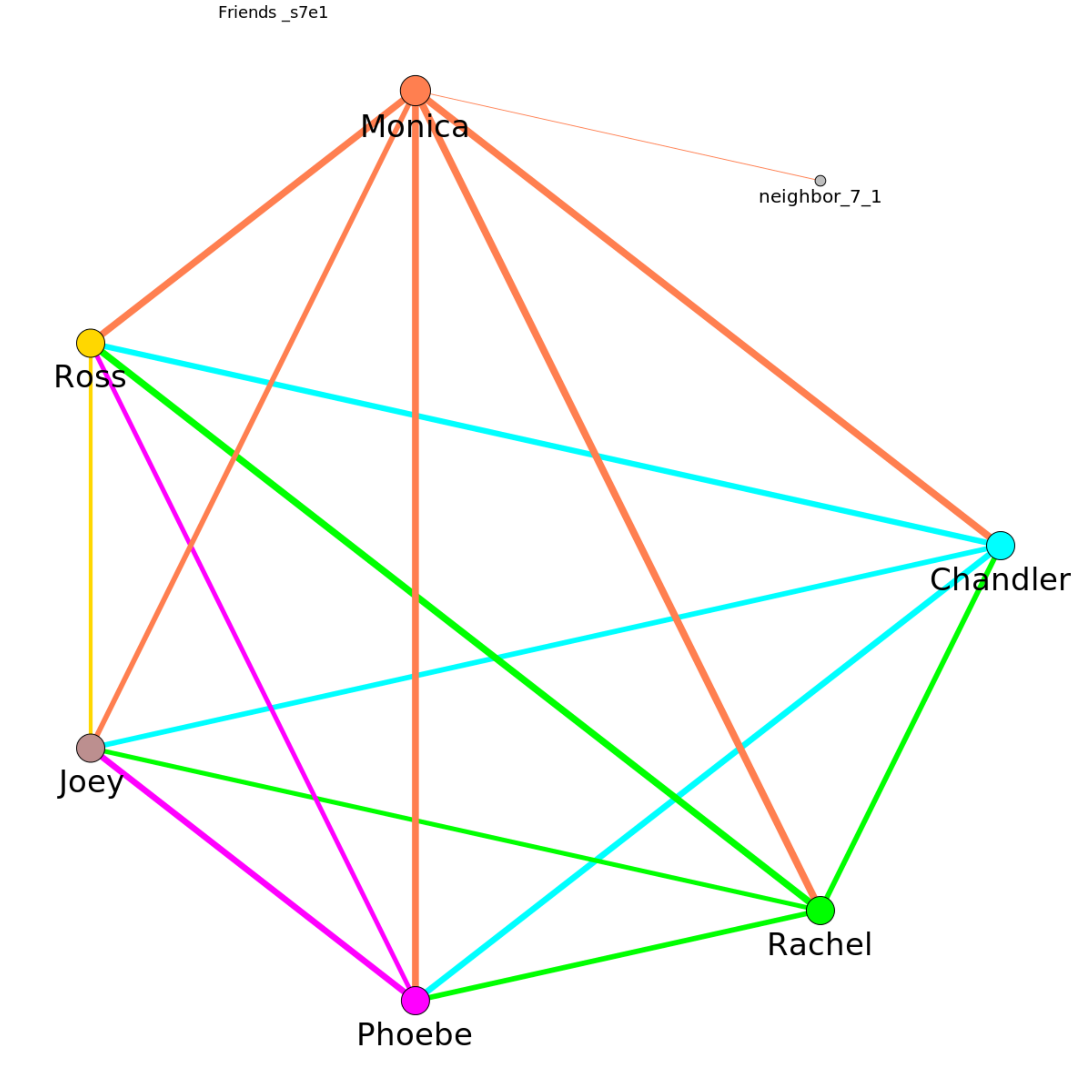}
 \caption{The six friends in a clique (edge's color refers to the node with higher number of connections)}
 \label{fig:clique}
\end{figure}

Are the six friends in fact equally central? 
What about point number 5 (M. Kauffmann's recollection about Chandler and Phoebe as ``a little more secondary'')?
Is it true that Monica is the mother hen,  ``the glue that holds this group together'' (as claimed in episode 3 in season 9, The One With The Pediatrician)?
Since Joey and Monica were, before the pilot, to form a romantic pair, do their interactions (especially in the beginning of the series) show that could have happened at all?
Do the six friends form a family?

The present paper uses \netth\ and visual tools to at least partially address these and other questions.
While this is spread throughout the text, the issue of family ties deserves some initial thoughts since it relates to the notion of \textbf{clique}.
In a graph $G=(V,E)$, where $V$ is the set of vertices (nodes) in the graph, and $E \subseteq V \times V $ is the set of edges,
clique is the subset of vertices in a graph, in which any two  vertices in the clique are directly connected.
The six friends form a clique in the majority of graphs.
One in particular deserves more attention because this graph refers to an episode that revolves practically only around the six friends: in the first episode of season 7 (The One with Monica's Thunder), the six form a clique as shown in Fig.~\ref{fig:clique}. Apart from their interactions,  there is only a brief utterance between  Monica and a  distant neighbor shouting at her.

\section*{Structure, centrality, and temporal aspects in the  social network of \fri}

The topological characterization of structures within  social networks is an  important step to recognize patterns of interest, and how they change along time.
For example, how distant are two vertices of the graph? How are these distances distributed regarding all pairs of vertices?
How to compare graphs that relate to different moments of a show?
How are the graphs partitioned?
Are there patterns that unveil when we see the plots and analyze their characteristic measures?

\begin{figure}[t]
 \centering
 \includegraphics[trim={0 0 0 2cm},clip,width=0.9\linewidth]{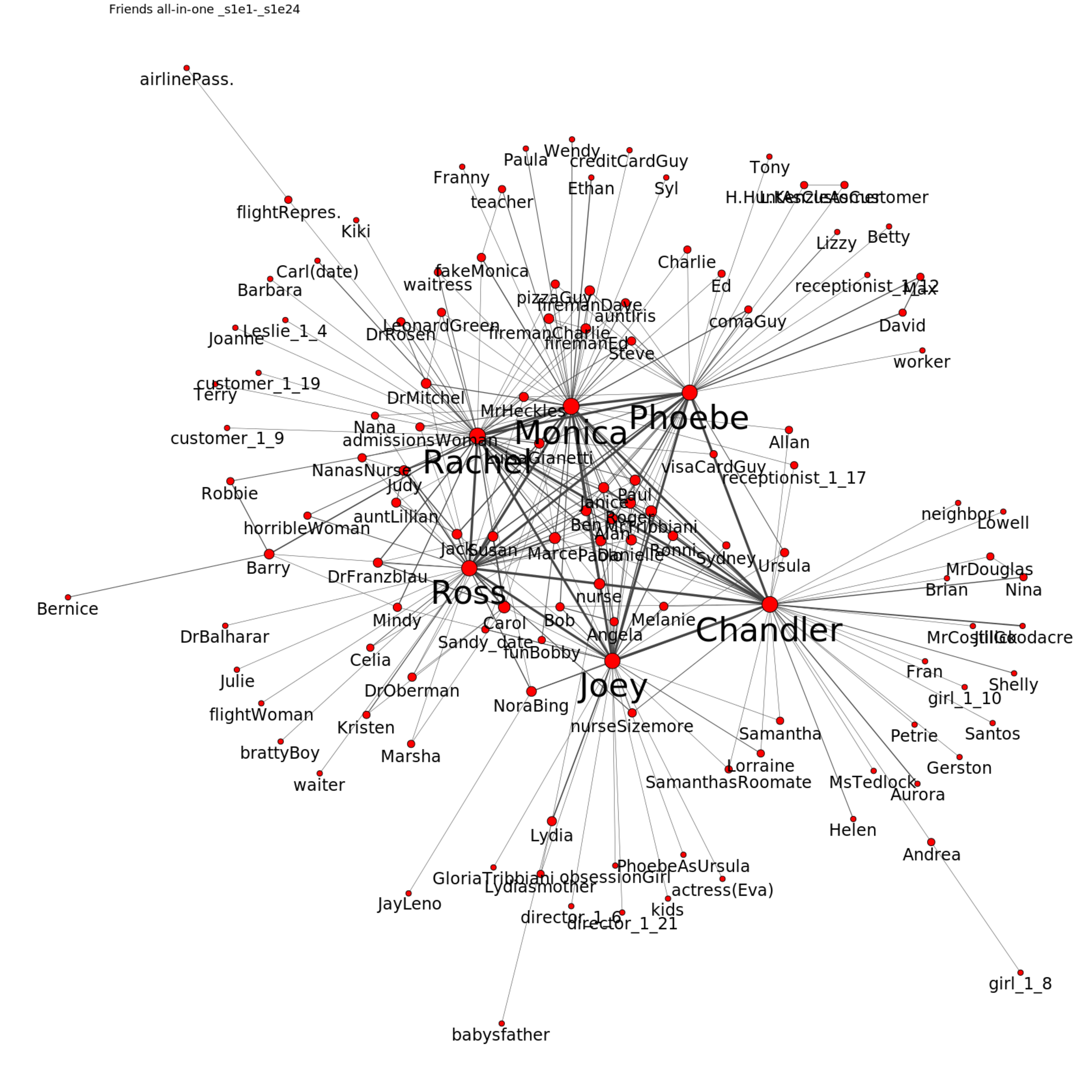}
 \caption{Graph corresponding to the entire first season
 }
 \label{fig:s1}
\end{figure}

\begin{figure}[t]
 \centering
 \includegraphics[trim={0 0 0 2cm},clip,width=0.9\linewidth]{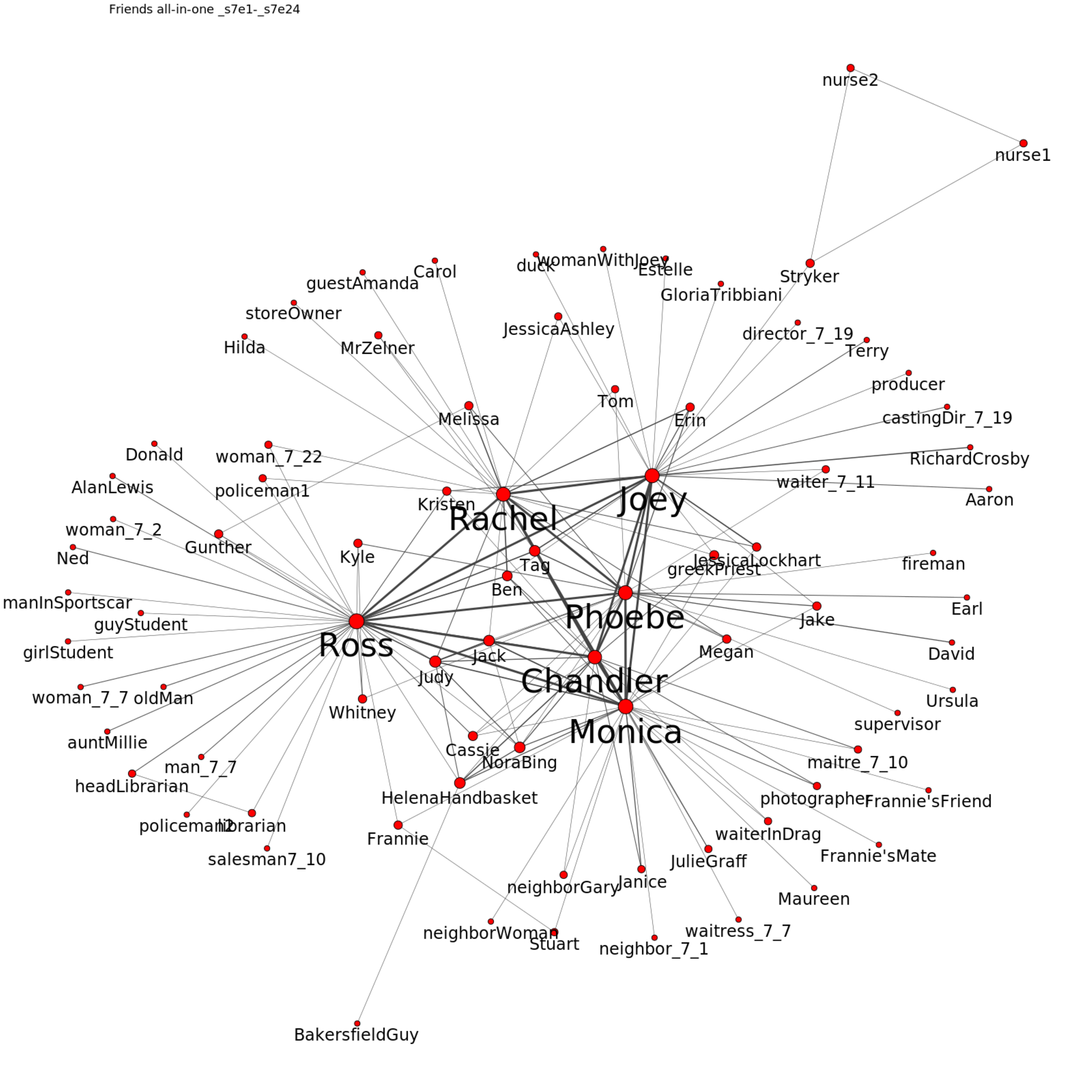}
 \caption{Graph corresponding to the entire season 7
}
 \label{fig:s7}
\end{figure}

For example, Fig.~\ref{fig:s1} and Fig.~\ref{fig:s7} show graphs that depict the interactions between all characters (vertices) that appear in season 1 (the one with the highest number of different vertices, $N=|V|=126$) and season 7 (the one with the lowest: $N=|V|=81$) respectively.
Not only these two graphs look different, with the former depicting, in the middle, more connections to vertices that represent other characters  (meaning that more characters interact with the six main characters); such difference can be numerically measured.

Take the clique of the graph
and the \textbf{clique number} (the cardinality of the clique).
In the graph of season 1 the clique number is 8. Obviously, the clique includes the six main characters. There are other two who interact among themselves and with the six, thus forming a clique.
In season 7, the clique number is just 7, meaning that only one character made to the group of six.

We might also be interested in knowing how likely two neighbors of a vertex  are to be connected, i.e., how likely it is that two friends of a character are also friends of each other. This is given by the graph's  clustering coefficient, that measures the ratio of connected triplets (3 vertices fully connected) by the ratio of possible triangles. 
In season 7'graph, the clustering coefficient is around 0.2. 
In season 1's, it is lower (0.15).
This relatively low ratios show that, along a season, lots of characters never get to meet others.
Of course, for individual episodes, clustering coefficients change a lot, with high values for episodes such as those that revolve mainly around the six friends (see later), in which this ratio is around 0.8, since basically everybody interacts with everybody throughout the whole episode.

Another difference between graphs in Fig.~\ref{fig:s1} and Fig.~\ref{fig:s7} relates to distances between vertices.
For this we resort to the concept of \textbf{geodesic} paths and \textbf{diameter} of a graph.
A geodesic  shortest path, between
vertices $i$ and $j$, is the one  with minimum length $d_{i,j}$ from vertex $i$ to vertex $j$.

To understand how this works mathematically speaking, here are some basic concepts related to how to measure such paths.
Be $d_{i,j}$  the length---number of edges along the path---of a geodesic path and $N=|V|$ the number of vertices in the graph\footnote{Henceforth, to follow the literature on networks, $N$ is used to denote $|V|$.}. 
In some cases we are interested in calculating the  mean geodesic distance for vertex $i$, which  is given by 
$\displaystyle{l_i = \frac{1}{N} \sum_j d_{i,j}}$; or we might simply look for the maximum geodesic path found in the graph, also known as diameter.

For the graphs of seasons 1 and 7,  diameters are 5 and 4 respectively.

These measures can be used to characterize any graph as shown for the examples in Fig.~\ref{fig:s1} and Fig.~\ref{fig:s7}.
Thus, they can be used to draw conclusions that refer to particular situations or time slices that appeared in the show.
Table~\ref{tab:situations} lists some situations of interest and characterize them in term of graph's size, diameter, clique number, and clustering. 
Notice that the number of edges does include multiple connections between some pairs of characters.
It starts with the whole graph, i.e., all episodes (henceforth AE), which aggregates the 236 graphs, one for each episode.
This graph has 746 vertices; the  diameter is 5, meaning that the two vertices  farthest apart  are only five hops from each other.
Notice that, in the whole table, diameters' values do not decrease much. 

Another two major aggregations of episodes (lines 2 and 3 in Table~\ref{tab:situations}) concern seasons 1 to 4 (s1--s4) and 5 to 10 (s5--s10) respectively.
The purpose of such aggregations is to investigate what happens with two of the friends, Monica and Chandler, who have started a relationship at the beginning of season 5.
Thus, it makes sense to analyze 
whether their interactions with other friends have been affected by their relationship.

Further, watching the show, it is possible to notice that generally, the first and last episodes of each season have different characteristics.
The first ones are focussed on recapping what was left in the air in the last episode of the previous season; they tend to be more focussed on the six friends and feature less characters.
Thus, it is interesting to compare the aggregated graphs of the first and last episodes (lines 4 and 5 in Table~\ref{tab:situations}).

Other three situations of interest are Thanksgiving episodes, episodes that involve flashbacks\footnote{Interactions in scenes from episodes with flashbacks were considered part of these episodes; thus some interactions that happened in other seasons can be found in these graphs.}, and episodes revolving around the six friends---normally with a single (or main) storyline\footnote{These are: s1e18 (The One with All the Poker), s2e3 (The One Where Heckles Dies), s3e2 (The One Where No One's Ready), s3e9 (The One with the Football), s3e16 (The One with the Morning After), s3e17 (The One Without the Ski Trip), s4e1 (The One with the Jellyfish), s4e12 (The One with the Embryos), s5e14 (The One Where Everybody Finds Out), s6e6 (The One on the Last Night), s6e9 (The One Where Ross Got High), s7e1 (The One with Monica's Thunder), s7e8 (The One Where Chandler Doesn't Like Dogs), s7e14 (The One Where They All Turn Thirty), s8e4 (The One with the Videotape), s8e9 (The One with the Rumor), s9e18 (The One with the Lottery), s10e4 (The One with the Cake), s10e10  (The One Where Chandler Gets Caught), s10e16 (The One with Rachel's Going Away Party).}.
Table~\ref{tab:situations} also lists the graphs for two single episodes (s1e1 and s10e18), as well as aggregated graphs corresponding to each of the ten seasons.

Comparing all  situations listed in Table~\ref{tab:situations}, we can see that, although the graphs have very different sizes,
the diameter does not vary much. 
It is possible to reach any two characters by traversing 5 edges at most.
Interestingly, this seem to corroborate the thesis of ``six degrees of separation" \citep{Karinthy1929,Milgram1967}, even for a fictional story.

The clustering coefficient tends to be very low for big graphs (because many characters never meet); they are  high for focussed episodes (Thanksgiving, those centered on the six friends, etc.), very high for episodes such as the pilot (s1e1); for graphs of each complete season clustering is around 0.2.

The highest clique number is  10, and it refers to the AE graph, i.e., the 236 episodes put together.
Recall that the show features around 750 characters over the ten seasons, thus one would expect  a higher clique number.
This supports  the claim by \cite{Marshall2007} that \fri\  is about a closed group:

``The six characters formed a culture
that no one else was allowed to enter. Even when Phoebe married Mike in season 10, Mike was
often working or had other commitments and, therefore, rarely interacted with the group.''


\begin{table}[t]
 \centering
 \caption{Graph's characteristics in different situations}
 \label{tab:situations}
 \begin{tabular}{c|ccccc} 
  \hline
  Graph & $N$ & $|E|$ & Diameter & Clique Nb. & Clustering Coef. \\ \hline
  all episodes (AE) & 746 & 16569 & 5 & 10 & 0.03 \\ 
  s1--s4 & 349 & 7675 & 5 & 8 & 0.07 \\
  s5--s10 & 462 & 8894 & 5 & 8 & 0.05 \\ \hline
  1st ep. all seasons & 43 & 749 & 3 & 7 & 0.34 \\
  last ep. all seasons & 69 & 690 & 5 & 7 & 0.30 \\ \hline
  Thanksgiving & 31 & 881 & 3 & 8 & 0.47 \\
  flashbacks & 69 & 1303 & 4 & 7 & 0.26 \\
  mainly the 6 & 49 & 2308 & 3 & 8 & 0.35 \\ \hline 
  s1e1 & 11 & 131 & 3 & 7 & 0.77 \\ 
  s10e18 & 19 & 79 & 4 & 6 & 0.47 \\ \hline 
  s1 & 126 & 2492 & 5 & 8 & 0.16 \\
  s2 & 107 & 1815 & 5 & 8 & 0.19 \\
  s3 & 98 & 1770 & 5 & 8 & 0.20 \\
  s4 & 96 & 1598 & 4 & 8 & 0.23 \\
  s5 & 93 & 1786 & 4 & 7 & 0.19 \\
  s6 & 99 & 1491 & 4 & 8 & 0.16 \\
  s7 & 81 & 1475 & 5 & 7 & 0.20 \\
  s8 & 110 & 1220 & 4 & 7 & 0.14 \\
  s9 & 101 & 1454 & 4 & 7 & 0.19 \\
  s10 & 87 & 1468 & 5 & 7 & 0.23 \\
  \hline 
 \end{tabular}
\end{table}

Measures such as clique number and diameter are meant to characterize a graph as a whole (see \cite{Costa+2007} for further details and definitions).
However, another method to characterize graphs takes the perspective of individual vertices (or edges) and analyzes how a given measure changes along those elements of the graph.
Centrality measures -- which apply, with small changes, to both vertices and edges -- are good examples.

\begin{figure}[t]
 \centering
 \includegraphics[trim={0 0 0 2cm},clip,width=0.7\linewidth]{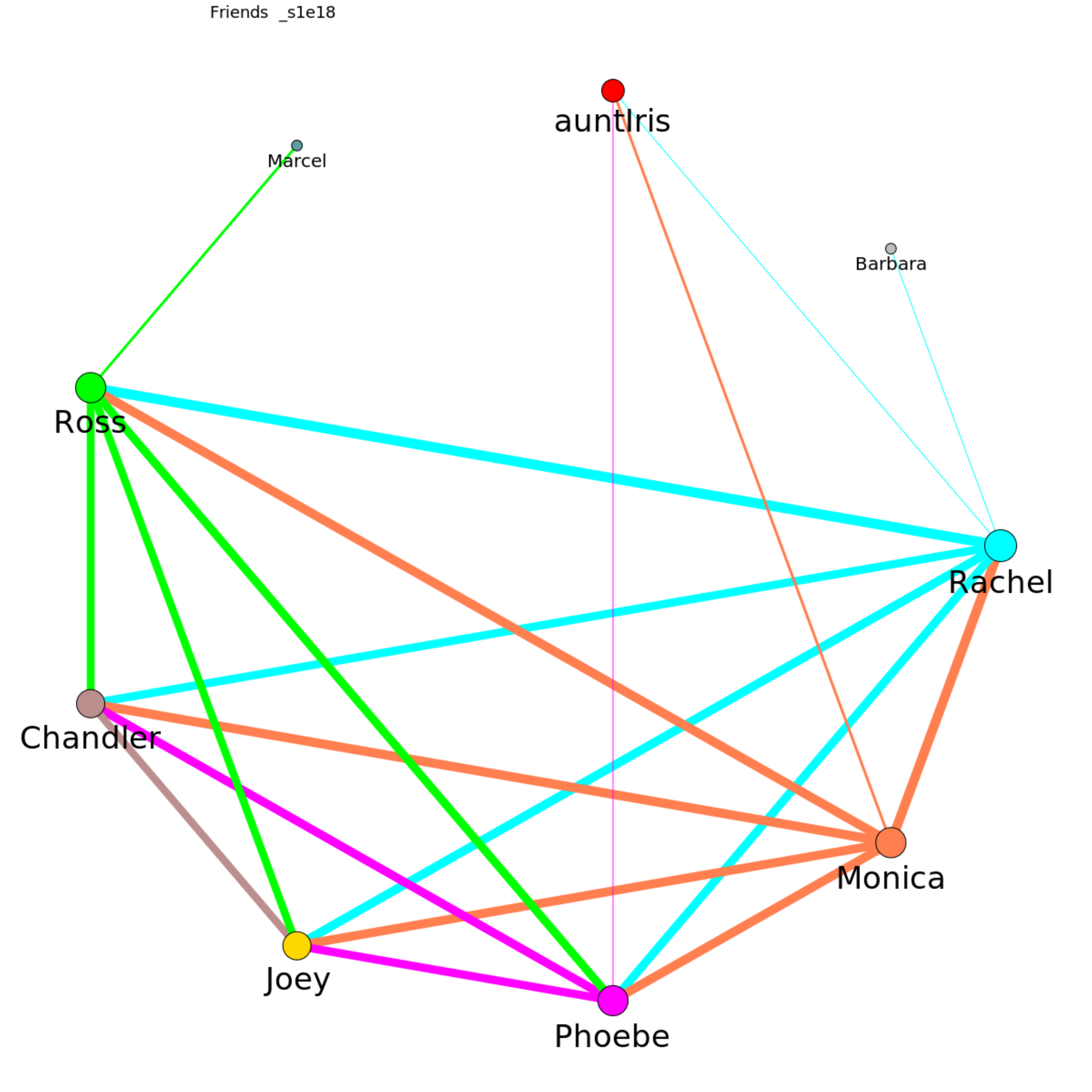}
 \caption{Graph corresponding to episode 18 of season 1 (The One with All the Poker): N=9; diameter=3; clustering=0.77}
 \label{fig:s1e18}
\end{figure}

\textbf{Degree} is the simplest centrality measure for a vertex $i$, corresponding to the number of connections  $i$ has, including multiple direct connections of $i$ to other vertices.
For example, in Fig.~\ref{fig:s1e18} (episode 18 in season 1) one can see that Rachel is directly connected to seven other characters. However, since she interacts several times with  other characters---especially with the other five friends---(this is represented in that figure by ticker edges), her  degree in that episode is 106.

\begin{figure}[t]
 \centering
 \includegraphics[trim={0 0 0 2cm},clip,width=0.9\linewidth]{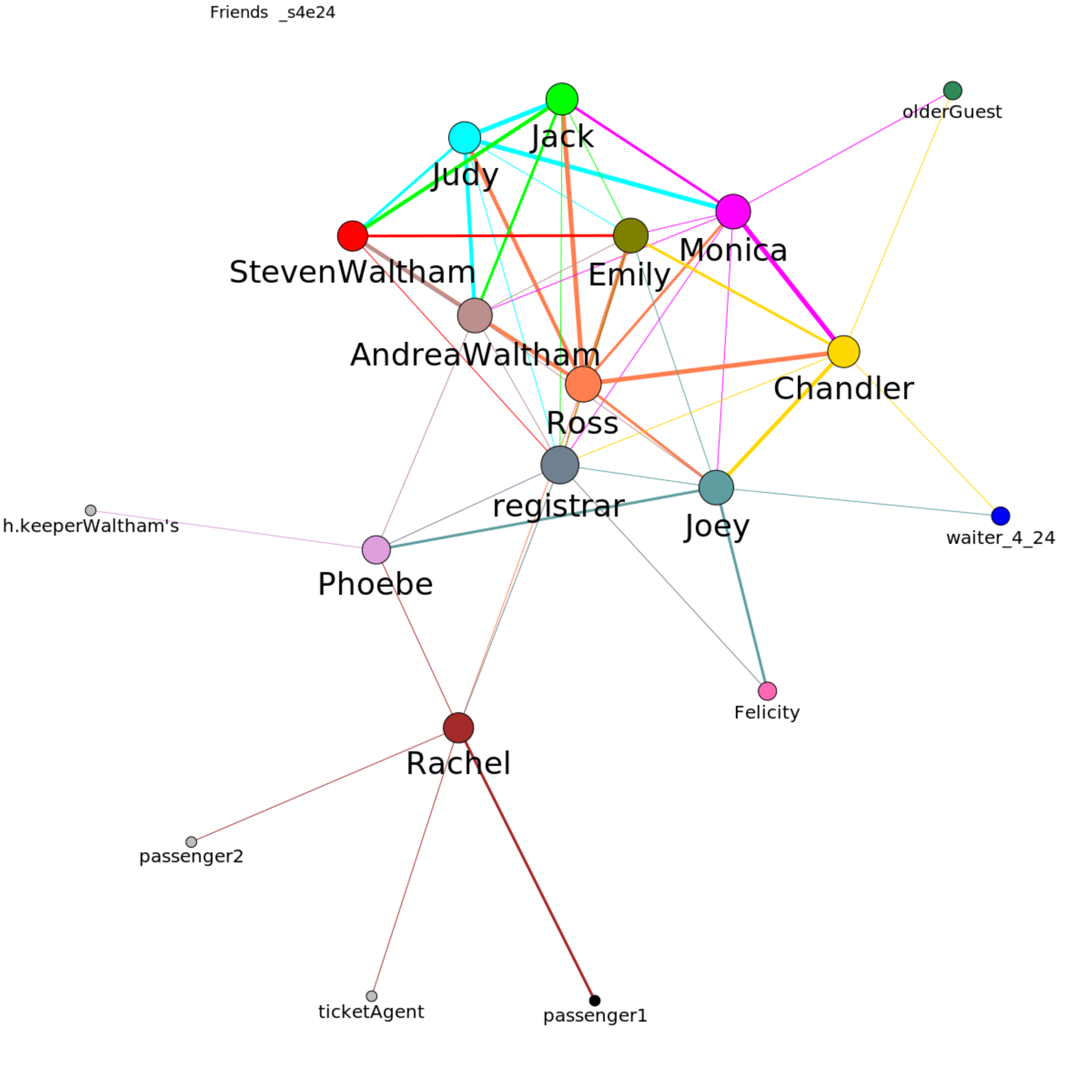}
 \caption{Graph corresponding to episode 24 of season 4 (The One with Ross's Wedding: Part Two): N=19; diameter = 4; clustering=0.65}
 \label{fig:s4e24}
\end{figure}

A vertex may have a low degree, be connected to other vertices that have likewise low degree, but still be  important in the network because it acts as a kind of bridge between groups of vertices.
For an example, see Fig.~\ref{fig:s4e24} (episode 24 in season 4) where the edges Ross--Rachel or Phoebe--Joey act as bridges between subgroups of characters.
Recall that, in this episode, while Ross was getting married in London, a pregnant Phoebe was stuck in New York trying to convince Rachel not to fly to London.
Since any path between two vertices that lie in different groups must pass through such  bridges, a measure called \textbf{betweenness} quantifies this kind of importance by measuring the extent to which a vertex lies on shortest paths between any two pairs of vertices.
    
Low values of this quantity mean that vertex $i$ is separated from other vertices by only a short distance (on average).
Thus, in a social network, this vertex would have easier access to information from or influence on other vertices.
This is important, e.g., for opinion dissemination, imitation, etc.

To compute the betweenness centrality,  let $n_{s,t}^{i}$ be the number of geodesic paths from $s$ to $t$ that pass through $i$ and let $n_{s,t}$ be the total number of geodesic paths from $s$ to $t$. 
Then the betweenness centrality of vertex $i$ is given by:

    $\displaystyle{b_i = \sum_{s, t} w_{s,t}^{i} = \sum_{s, t} \frac{n_{s,t}^{i}}{n_{s,t}}}$ 

If  $n_{s,t} = 0$, then $w_{s,t}^{i} = 0$.
This quantity can be  rescaled by dividing it by the number of pairs of nodes not including $i$ so that $b_i \in [ 0 , 1 ]$, which is $( N - 1 ) ( N - 2 ) /2$ (for undirected graphs).

Note that the previously defined mean distance $l_i$ of more central vertices correspond to low values, whereas high values correspond to less central ones.
Thus, this acts in the opposite way as defined for other centrality measures such as degree, for instance.
The inverse of the mean distance $l_i$ is called  \textbf{closeness} centrality: 

    $\displaystyle{C_i = \frac{1}{l_i} = \frac{N}{\sum_j d_{i,j}}}$

\section*{How centrality changes 
in \fri}

\definecolor{s1}{RGB}{228, 26, 28}
\definecolor{ch}{RGB}{55, 126, 184}
\definecolor{jo}{rgb}{0.13, 0.7, 0.67}
\definecolor{ro}{RGB}{152, 78, 163}
\definecolor{s5}{RGB}{255, 127, 0}
\definecolor{ra}{rgb}{0.89, 0.44, 0.48}
   
\pgfplotstableread[row sep=\\,col sep=comma]{
cm, Monica, Chandler, Ross, Rachel, Joey, Phoebe\\
S1E1, 55, 45, 33, 46, 40, 24 \\     
S10E18, 25, 24, 21, 15, 22, 26 \\
}\nonnormdegrees

\pgfplotsset{compat=1.11,
    /pgfplots/ybar legend/.style={
    /pgfplots/legend image code/.code={%
       \draw[##1,/tikz/.cd,yshift=-0.25em]
        (0cm,0cm) rectangle (3pt,0.8em);},
   },
}

\begin{figure}[t]
 \centering
\begin{tikzpicture}
\begin{axis}[
  ybar,
  bar width=12pt,
  enlarge x limits={0.5},  
  ymin = 10,
  ymax = 60,
  symbolic x coords={S1E1,S10E18},
  xtick = data,
  nodes near coords,
  nodes near coords  style={font=\tiny},  
  legend style={at={(0.5,-0.25)},
  anchor=north,legend columns=-1},
  width=0.8\textwidth,
  height=0.4\textwidth,
  ]
\addplot [fill=red] table [x=cm,y=Monica]{\nonnormdegrees};
\addplot [fill=ch] table [x=cm,y=Chandler]{\nonnormdegrees};
\addplot [fill=ro] table [x=cm,y=Ross]{\nonnormdegrees};
\addplot [fill=ra] table [x=cm,y=Rachel]{\nonnormdegrees};
\addplot [fill=jo] table [x=cm,y=Joey]{\nonnormdegrees};
\addplot [fill=orange] table [x=cm,y=Phoebe]{\nonnormdegrees};
\legend{Monica,Chandler,Ross,Rachel,Joey,Phoebe}
\end{axis}
\end{tikzpicture}
 \caption{Degrees in S1E1 and S10E18.}
 \label{fig:nonnormdegreebar}
\end{figure}
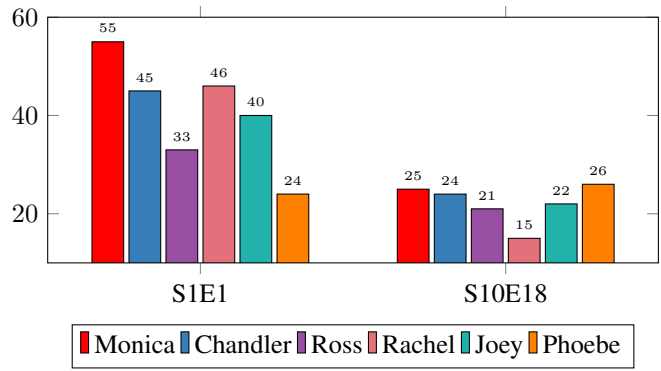

\pgfplotsset{every axis/.append style={
                    label style={font=\small},
                    tick label style={font=\scriptsize}  
                    }}
                    
\pgfplotstableread[row sep=\\,col sep=comma]{
cm, Monica, Chandler, Ross, Rachel, Joey, Phoebe\\
S1, 6.46,6.08,6.04,5.84,5.18,5.17\\
S2, 5.15,5.11,5.24,4.42,4.48,4.39\\
S3, 5.32, 5.60, 5.38,4.81,5.38,4.98\\
S4,4.69,5.41,4.56,4.76,4.48,4.56\\
S5,5.93,6.10,5.76,5.03,6.00,5.64\\
S6,4.29,4.71,4.20,4.35,4.77,4.00\\
S7,6.41,5.56,5.18,5.43,5.05,4.99\\
S8,3.16,2.66,3.28,3.35,3.33,2.72\\
S9,4.08,4.09,4.04,3.81,3.74,3.75\\
S10,5.13,5.31,4.49,4.48,4.64,4.45\\
}\normdegrees

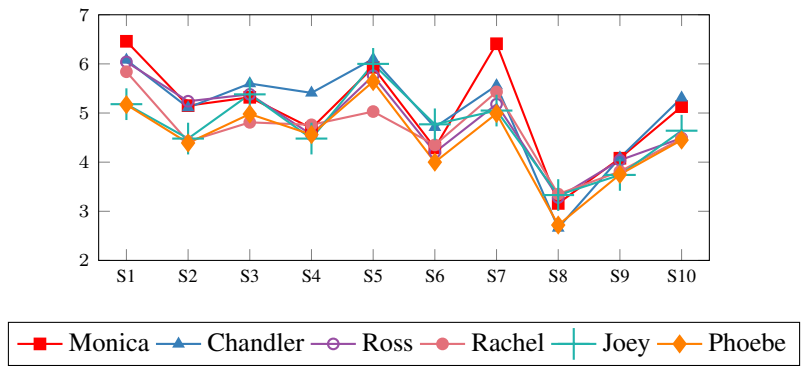
\begin{figure}[t]
 \centering
\begin{tikzpicture}
\begin{axis}[
  enlarge x limits={0.05},  
  ymin = 2,
  ymax = 7,
  ytick distance=1,
  symbolic x coords={S1,S2,S3,S4,S5,S6,S7,S8,S9,S10},
  xtick = data,
  legend style={at={(0.5,-0.25)},
  anchor=north,legend columns=-1},
  every axis plot/.append style={thick},
  width=0.8\textwidth,
  height=0.4\textwidth,
  ]
\addplot [color=red,mark=square*] table [x=cm,y=Monica]{\normdegrees};
\addplot [color=ch,mark=triangle*] table [x=cm,y=Chandler]{\normdegrees};
\addplot [color=ro,mark=o] table [x=cm,y=Ross]{\normdegrees};
\addplot [color=ra,mark=*] table [x=cm,y=Rachel]{\normdegrees};
\addplot [color=jo,mark=+, mark size=6pt] table [x=cm,y=Joey]{\normdegrees};
\addplot [color=orange,mark=diamond*, mark size=3pt] table [x=cm,y=Phoebe]{\normdegrees};
 \legend{Monica,Chandler,Ross,Rachel,Joey,Phoebe}
\end{axis}
\end{tikzpicture}
 \caption{Normalized degrees through all seasons.}
 \label{fig:degs1s10}
\end{figure}


   
\pgfplotstableread[row sep=\\,col sep=comma]{
cm, Monica, Chandler, Ross, Rachel, Joey, Phoebe\\
S1-S4,6.67,6.79,6.53,6.10,5.97,5.84\\
S5-S10,5.80,5.69,5.43,5.32,5.55,5.13\\
S1E1,5.50,4.50,3.30,4.60,4.00,2.40\\
S10E18,1.32,1.26,1.11,0.79,1.16,1.37\\
1st,5.98,5.38,5.81,5.62,4.95,4.26\\
last,2.58,2.59,2.70,2.70,2.46,2.49\\
thanksgiving,10.13,9.87,8.87,8.43,8.20,6.93\\
the 6, 16.72,15.91,15.47,15.40,14.77,14.09\\
AE,6.69,6.69,6.40,6.14,6.21,5.90\\
}\ndegdiffsit
             
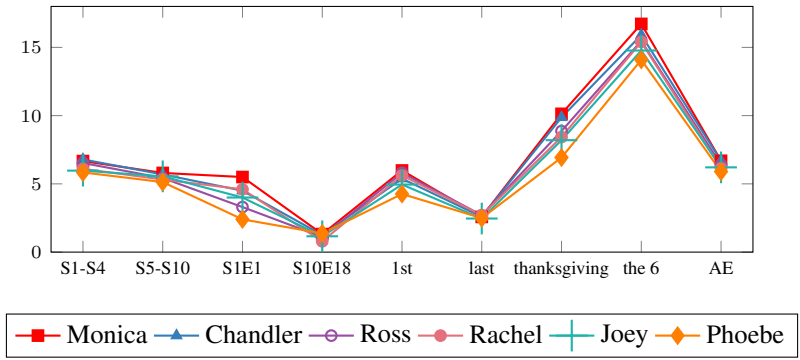
\begin{figure}[t]
 \centering
\begin{tikzpicture}
\begin{axis}[
  enlarge x limits={0.05},  
  ymin = 0,
  ymax = 18,
  symbolic x coords={S1-S4,S5-S10,S1E1,S10E18,1st,%
  last,thanksgiving,the 6,AE},
  xtick = data,
  legend style={at={(0.5,-0.25)},
  anchor=north,legend columns=-1},
  width=0.9\textwidth,
  height=0.4\textwidth,
  every axis plot/.append style={thick},
  ]
\addplot [color=red,mark=square*] table [x=cm,y=Monica]{\ndegdiffsit};
\addplot [color=ch,mark=triangle*] table [x=cm,y=Chandler]{\ndegdiffsit};
\addplot [color=ro,mark=o] table [x=cm,y=Ross]{\ndegdiffsit};
\addplot [color=ra,mark=*] table [x=cm,y=Rachel]{\ndegdiffsit};
\addplot [color=jo,mark=+, mark size=6pt] table [x=cm,y=Joey]{\ndegdiffsit};
\addplot [color=orange,mark=diamond*, mark size=3pt] table [x=cm,y=Phoebe]{\ndegdiffsit};
 \legend{Monica,Chandler,Ross,Rachel,Joey,Phoebe}
\end{axis}
\end{tikzpicture}

 \caption{Normalized degrees for different situations (1st and last mean aggregation of all first and last episodes; the 6 means episodes where mainly the six friends interact; AE are all episodes).}
 \label{fig:normdegreeline}
\end{figure}

In order to answer those questions mentioned before about the known facts of \fri, one can look at centrality measures of  characters.
How do their degrees compare and change with time or with situations (as listed in Table~\ref{tab:situations})?
Such comparison requires that degree is normalized to account for the graph's size $N$.
Here is why: Fig.~\ref{fig:nonnormdegreebar} shows degrees for the six friends in episode 1 of season 1 (The  One Where Monica Gets a Roommate) and episode 18 in season 10 (The Last One: Part 2), where $N$ is 11 and 20 respectively. 
Although the former graph has less vertices, the values of degrees are higher.
Thus, we cannot compare them directly; it is important to account for the size of the graph.
Normalized degree  is  obtained by dividing the degree by $N-1$. 

Henceforth, degree values appear as normalized quantities.
Such normalization is even more necessary when comparing graphs that aggregate different number of episodes, as some listed in Table~\ref{tab:situations}.

Fig.~\ref{fig:degs1s10} shows normalized degree for the six friends through seasons 1 to 10, while
Fig.~\ref{fig:normdegreeline} shows  some other situations of interest.

Let us first take a look at degree values throughout seasons and situations and compare some of  them.
Considering all seasons (Fig.~\ref{fig:degs1s10}), there is a tendency of alternating high degrees (odd seasons) with lower ones (even seasons), with a noticeable low in season 8, where all six friends have their lowest degree. They then increase for seasons 9 and 10.

Comparing situations (Fig.~\ref{fig:normdegreeline}), degrees for the last episodes of each season are  much lower than those for the first episodes of the seasons, probably due to the fact that, although there are less characters in the first episodes, more interactions happen among these characters; thus there are more  edges in the  graph.
Seen another way, it may be that scenes take longer in the last episodes of each season, thus there are fewer interactions (that take longer though).
The same is true when we compare the very first (s1e1) with the very last (s10e18) episodes.

The highest degree values are associated with episodes revolving around the six friends, probably because the six interact a lot (even if the size of the graph is small).
In Thanksgiving episodes,  degrees are also high.
In both these two situations, degrees are higher than when one considers the average degree over the 236 episodes (Table~\ref{tab:236ep}), especially for Monica. 

\begin{table}[t]
 \centering
 \caption{Centrality measures (all 236 episodes)}
 \begin{tabular}{cccccccccccccccccccc}
 \hline
 Centrality & Monica & Chandler & Ross & Rachel & Joey & Phoebe & 7th \\ \hline
 degree & 6.70	&	6.70	&	6.40	&	6.15	&	6.22	&	5.90 & 0.24 \\
 closeness & 	0.57	& 0.58	&	0.59	&	0.58	&	0.60	&	0.57 & 0.49\\
 betweenness &	0.19	&	0.24	&	0.29	&	0.25	&	0.33	&	0.20 & $\approx$ 0 \\
 \hline
 \end{tabular}
 \label{tab:236ep}
\end{table}

Now, what happens when we compare degrees of the six friends within the same situation? 
With few exceptions (notably Chandler in season 4, Monica in season 7, in s1e1 and in Thanksgiving episodes), we can conclude that there is little difference between degrees of the six friends in each situation (though, as just discussed, degree values vary from situation to situation), and that, normally, Phoebe's is among the lowest degrees, whereas Monica's is among the highest.
Line 1 (degree) in Table~\ref{tab:236ep} confirms this by detailing what happens regarding the AE graph: when the whole show is considered, there is little difference regarding the values of degree of the six friends.

For sake of comparison, still in the AE graph, the seventh character in the ranking in terms of degree (actually Mike, Judy, and Jack have roughly the same value) is also included in Table~\ref{tab:236ep}. Degree's value for this character is just a small fraction of the others, thus, again supporting the thesis of a closed group.
On a related note, the degree assortativity---preference for a network's nodes to attach to others that have similar degree---is low in the AE graph: just 0.023.

Now, what happens in terms of closeness and betweenness?
As for the former, Table~\ref{tab:236ep} shows that this value does not change much from friend to friend, and contrarily to what happens with degrees, also characters outside the group of six have relatively high closeness.
This happens because the distribution of distances ($l_i$ as defined before) is relatively uniform since virtually all characters are connected to the clique.
By the way, another popular centrality measure, the Eigenvector centrality, follows a similar pattern for the same reason.

However, the most interesting characteristic of the six friends is their normalized betweenness.
Starting with the AE graph (Table~\ref{tab:236ep}), we see that there is a difference between the values, with Joey having the highest (0.33), while Monica and Phoebe have the lowest.
This means that Joey (an Italo-American!) is the guy who more efficiently connects a lot of other characters, while Monica, who has the highest degree, is not an impressing connector.
This might point to she being the queen in her own apartment.

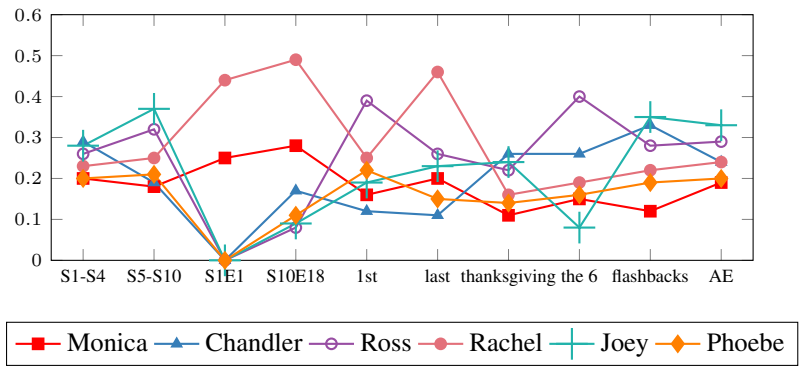
\begin{figure}[t]
 \centering
\begin{tikzpicture}
  \begin{axis}[
  enlarge x limits={0.05}, 
  ymin = 0.,
  ymax = 0.6,
  ytick distance=0.1,
  symbolic x coords={S1-S4,S5-S10,S1E1,S10E18,1st,%
  last,thanksgiving,the 6,flashbacks,AE},
  xtick = data,
  legend style={at={(0.5,-0.25)},
  anchor=north,legend columns=-1},
  width=0.9\textwidth,
  height=0.4\textwidth,
  every axis plot/.append style={thick},
  ]
\addplot [color=red,mark=square*] table [x=cm,y=Monica]{\normbetdiffsit};
\addplot [color=ch,mark=triangle*] table [x=cm,y=Chandler]{\normbetdiffsit};
\addplot [color=ro,mark=o] table [x=cm,y=Ross]{\normbetdiffsit};
\addplot [color=ra,mark=*] table [x=cm,y=Rachel]{\normbetdiffsit};
\addplot [color=jo,mark=+, mark size=6pt] table [x=cm,y=Joey]{\normbetdiffsit};
\addplot [color=orange,mark=diamond*, mark size=3pt] table [x=cm,y=Phoebe]{\normbetdiffsit};
 \legend{Monica,Chandler,Ross,Rachel,Joey,Phoebe}
\end{axis}
\end{tikzpicture}
 \caption{Normalized Betweenness different situations}
 \label{fig:betSituations}
\end{figure}

\begin{figure}
 \centering
\begin{tikzpicture}
  \begin{axis}[
  enlarge x limits={0.05}, 
  ymin = 0.,
  ymax = 0.6,
  ytick distance=0.1,
  symbolic x coords={S1,S2,S3,S4,S5,S6,S7,S8,S9,S10},
  xtick = data,
  legend style={at={(0.5,-0.25)},
  anchor=north,legend columns=-1},
  width=0.9\textwidth,
  height=0.4\textwidth,
  every axis plot/.append style={thick},
  ]
\addplot [color=red,mark=square*] table [x=cm,y=Monica]{\normbet};
\addplot [color=ch,mark=triangle*] table [x=cm,y=Chandler]{\normbet};
\addplot [color=ro,mark=o] table [x=cm,y=Ross]{\normbet};
\addplot [color=ra,mark=*] table [x=cm,y=Rachel]{\normbet};
\addplot [color=jo,mark=+, mark size=6pt] table [x=cm,y=Joey]{\normbet};
\addplot [color=orange,mark=diamond*, mark size=3pt] table [x=cm,y=Phoebe]{\normbet};
 \legend{Monica,Chandler,Ross,Rachel,Joey,Phoebe}
\end{axis}
\end{tikzpicture}
 \caption{Normalized betweeeness seasons 1 to 10}
 \label{fig:bets1s10}
\end{figure}
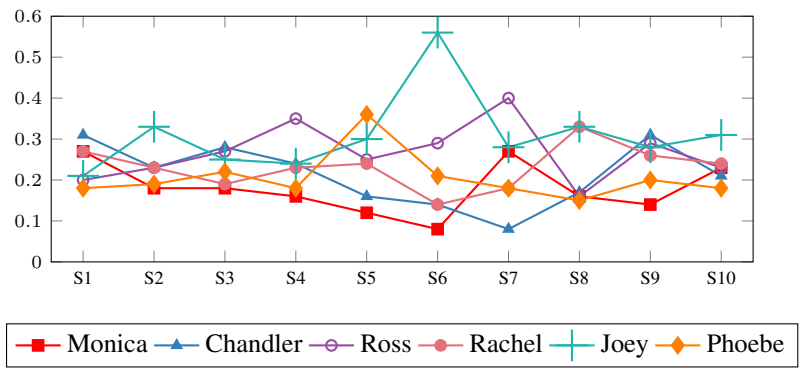

Unlike degree 
and  closeness, betweenness of the six friends varies a lot within a situation or temporal slice, as depicted in Fig.~\ref{fig:betSituations} and Fig.~\ref{fig:bets1s10}.
In the pilot (s1e1), only Rachel and Monica have non-zero betweenness.
Rachel also has the highest value in the final episode (s10e18), in which there is a disparity in the betweenness of the six friends:
Ross and Phoebe, who are too busy chasing Rachel, have the lowest values.
Comparing the first episodes in each season to the last ones, we see that Ross dominates the former situation, while Rachel dominates in the latter.
In fact, Rachel has a very prominent role in many last episodes: in season 1 (The One Where Rachel Finds Out), season 2 (The One with Barry and Mindy's Wedding),  season 4 (The One with Ross's Wedding: Part Two), season 5 (The One in Vegas: Part 2), season 8 (The One Where Rachel Has a Baby: Part 2), and season 10.

Because all Thanksgiving dinners take place at Monica's, she has high degree; however since she is probably too busy cooking, her betweenness is low as she has no time to connect with further characters.

The show also explores flashbacks: there are some episodes where the six friends reminisce about past experiences.
A graph that includes such episodes shows that Joey and Chandler dominate the flashbacks in terms of betweenness, i.e., they connect more characters who appear in these episodes.

Throughout the seasons, betweenness values also changes a lot (Fig.~\ref{fig:bets1s10}), with a very noticeable role of Joey in season 6 (who by the way has a high value in general).
Notice that Monica's betweeeness centrality decreases season after season until she  marries Chandler in season 7 (her highest value).
Chandler, on the contrary, has his lowest betweenness exactly in this same season!

Apart from the six friends, as said, Mike, Judy, and Jack have the highest degree values.
However, Mike is not a good connector. His betweenness is nearly half of that of Richard, Joshua, Pete, Dr. Hobart, and even a waiter.
In season 5, Frank Jr. has the highest betweenness after the six friends despite having only the 10th highest degree.
Conversely, in season 7, Tag has the 7th highest degree but very low betweenness.

\pgfplotstableread[row sep=\\,col sep=comma]{
cm, Monica, Chandler, Ross, Rachel, Joey, Phoebe\\
S1-S4,	0.20,	0.29,	0.26,	0.23,	0.28,	0.20\\
S5-S10,	0.18,	0.19,	0.32,	0.25,	0.37,	0.21\\
S1E1,	0.25,	0.00,	0.00,	0.44,	0.00,	0.00\\
S10E18,	0.28,	0.17,	0.08,	0.49,	0.09,	0.11\\
1st,	0.16,	0.12,	0.39,	0.25,	0.19,	0.22\\
last,	0.20,	0.11,	0.26,	0.46,	0.23,	0.15\\
thanksgiving,	0.11,	0.26,	0.22,	0.16,	0.24,	0.14\\
the 6,	0.15,	0.26,	0.40,	0.19,	0.08,	0.16\\
flashbacks,	0.12,	0.33,	0.28,	0.22,	0.35,	0.19\\
AE,	0.19,	0.24,	0.29,	0.24,	0.33,	0.20\\
}\normbetdiffsit

\pgfplotstableread[row sep=\\,col sep=comma]{
cm, Monica, Chandler, Ross, Rachel, Joey, Phoebe\\
S1, 0.27,0.31,	0.20,	0.27,	0.21,	0.18\\
S2,	0.18,	0.23,	0.23,	0.23,	0.33,	0.19\\
S3,	0.18,	0.28,	0.27,	0.19,	0.25,	0.22\\
S4,	0.16,	0.24,	0.35,	0.23,	0.24,	0.18\\
S5,	0.12,	0.16,	0.25,	0.24,	0.30,	0.36\\
S6,	0.08,	0.14,	0.29,	0.14,	0.56,	0.21\\
S7,	0.27,	0.08,	0.40,	0.18,	0.28,	0.18\\
S8,	0.16,	0.17,	0.16,	0.33,	0.33,	0.15\\
S9,	0.14,	0.31,	0.29,	0.26,	0.28,	0.20\\
S10,	0.23,	0.21,	0.23,	0.24,	0.31,	0.18\\
}\normbet

\subsection*{No dominant storyline?}

Recall that one assumption about \fri\ is that stories have (nearly) equal weight within an episode.
In fact, according to E. \cite{Kolbert1994}:

``One change NBC requested but the writers refused to make was to play up one story line and play down the others. (Most television comedy episodes are written with a primary story, the A story, and a secondary story, the B story.) Instead, they kept three story lines of essentially equal weight.''

However,  \cite{Thompson2003} argues that this was not the case in terms of screen time (pages 56--57).
To make her point, \cite{Thompson2003} describes what happens in episode 11 of season 7 (The One with All the Cheesecakes), where there were
three stories, each involving two of the characters.
She concludes that ``the three plotlines were not accorded equal time''. One (Phoebe and Joey commitment to dinner) ``was
given distinctly more weight''
(nearly half the total screen time, despite the title of the episode, which involved Chandler and Rachel's obsession with cheesecakes).
The third was about Monica and Ross attending their cousin's wedding.

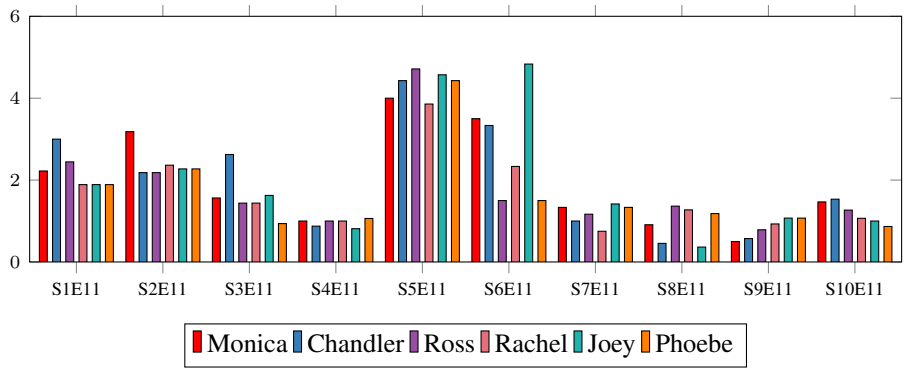
\begin{figure}[ht]
\begin{tikzpicture}
\begin{axis}[
  ybar,
  x=1.15cm,
  bar width=3pt,
  enlarge x limits={0.06},  
  ymin = 0,
  ymax = 6,
  symbolic x coords={S1E11,S2E11,S3E11,S4E11,S5E11,S6E11,S7E11,S8E11,S9E11,S10E11},
  xtick = data,
  legend style={at={(0.5,-0.25)},
  anchor=north,legend columns=-1},
  width=1.0\textwidth,
  height=0.4\textwidth,
  ]
\addplot [fill=red] table [x=cm,y=Monica]{\normdegepeleven};
\addplot [fill=ch] table [x=cm,y=Chandler]{\normdegepeleven};
\addplot [fill=ro] table [x=cm,y=Ross]{\normdegepeleven};
\addplot [fill=ra] table [x=cm,y=Rachel]{\normdegepeleven};
\addplot [fill=jo] table [x=cm,y=Joey]{\normdegepeleven};
\addplot [fill=orange] table [x=cm,y=Phoebe]{\normdegepeleven};
 \legend{Monica,Chandler,Ross,Rachel,Joey,Phoebe}
\end{axis}
\end{tikzpicture}

 \caption{Normalized degree for  episode number 11 of each season.}
 \label{fig:d11}
\end{figure}

In the present paper  degree centrality is used to shed light on this issue.
Degree, albeit not fully, is a reasonable measure of the dominance of a character.
Assuming that this dominance reflects the dominance of the storyline she or he is involved, can  degree values of the six friends say something about the issue of dominance of a storyline?
Fig.\ref{fig:d11} shows degree values for the six friends in several episodes,  including the  cheesecake episode (s7e11).
In this episode, Joey and Phoebe indeed have  high degrees, but so does Monica.
Adding Joey and Phoebe's as well as Ross and Monica's degrees, these sums are nearly the same, but higher than Chandler plus Rachel's.
This seems to confirm \cite{Thompson2003}'s point that ``it is implausible that [in sitcoms] each story thread is a main story in its own right".

What about the other episodes? Due to the impossibility to show and analyze each  of them, this paper concentrates on a subset.
To continue the analysis of \cite{Thompson2003}, the 11th episode in each season was selected (Fig.\ref{fig:d11}).

Joey's degree stands out in s6e11 (The One with the Apothecary Table), where Joey is caught in the middle when Janine tells him she doesn't like Monica and Chandler. 
Interestingly, again, the title of the episode is about characters with low degree: Rachel and Ross both buy  apothecary tables from a store that Phoebe hates.

Analyzing degree centrality of the six characters, it is hard to say that there  is a balance in story threads in single episodes as those in Fig.\ref{fig:d11}.

\pgfplotstableread[row sep=\\,col sep=comma]{
cm, Monica, Chandler, Ross, Rachel, Joey, Phoebe\\
S1E11,  2.22222222222,  3., 2.44444444444,  1.88888888889,  1.88888888889,  1.88888888889\\
S2E11,  3.18181818182,  2.18181818182,  2.18181818182,  2.36363636364,  2.27272727273,  2.27272727273\\
S3E11,  1.5625, 2.625,  1.4375, 1.4375, 1.625,  0.9375\\
S4E11,  1., 0.875,  1., 1., 0.8125, 1.0625\\
S5E11,  4., 4.42857142857,  4.71428571429,  3.85714285714,  4.57142857143,  4.42857142857\\
S6E11,  3.5,    3.33333333333,  1.5,  2.33333333333,  4.83333333333,  1.5\\
S7E11,  1.33333333333,  1., 1.16666666667,  0.75, 1.41666666667,  1.33333333333\\
S8E11,  0.909090909091, 0.454545454545, 1.36363636364,  1.27272727273,  0.363636363636, 1.18181818182\\
S9E11,  0.5,  0.571428571429, 0.785714285714, 0.928571428571, 1.07142857143,  1.07142857143\\
S10E11, 1.46666666667,  1.53333333333,  1.26666666667,  1.06666666667,  1., 0.866666666667\\
}\normdegepeleven

\vspace{0.5cm}
\subsection*{Pairwise connections and surrogate family}
Centrality measures say nearly nothing about how any two characters interact. 
Therefore, from three of the aggregated graphs already discussed, the number of interactions per pair of characters were recorded, i.e., the number of edges between all pairs formed by permutation of the six friends.
These three graphs are: AE, s1--s4, and s5--s10.

Considering the AE, the highest number of interactions occurred between Chandler and Monica (1139 times), followed by Chandler and Joey (1040 times), Ross and Rachel (945), and Monica and Phoebe (915).
In fact, Monica interacts more with   Phoebe than with Rachel (820 times), which is surprising giving that they were high-school best friends as well as room mates for five seasons.
Chandler and Rachel interact only 615 times (in fact there is only one episode in which they have a more consistent interaction--- The One with All the Cheesecakes); next come Ross and Phoebe (661). 
Despite the family relationship, Ross and Monica  (brothers) interact only 736 times (noteworthy: in seasons 5 to 10, Ross is the least interacting character to Monica). 
As a comparison, Ross interacts with Chandler 821 times.
Monica and Joey interact only 752 times (in fact, in seasons 1 to 4, Joey was the least interacting character to Monica),  which is a good indication that the idea of them forming the romantic pair was abandoned for good.

Apart from interactions among the six main characters, the highest number of interactions (still, much lower than those just mentioned) happen between Phoebe and Mike (69) and Chandler and Janice (48).
Only then come the number of interactions among birth family members (Ross and Monica to Jack and Judy, between 40 and 47), which seems to support the concept of surrogate family ties in \fri.

\begin{table}[t]
 \centering
 \caption{Number of interactions per season between any two of the main characters; blue refer to seasons 1 to 4  and magenta to seasons 5 to 10; bold and italics refer to high and low figures respectively.}
 \label{tab:interactions}
 \begin{tabular}{c|cccccc}
  & Monica & Chandler & Ross & Rachel & Joey & Phoebe \\ \hline
  Monica & -- & \color{blue}101 & \color{blue}96 & \color{blue}106 &  \color{blue}89 & \color{blue}111 \\
  Chandler & \textbf{\color{magenta}123} & -- & \color{blue}102 & \color{blue}81 & \textbf{\color{blue}132} & \color{blue}92			\\ 
  Ross & \textit{\color{magenta}59} & \color{magenta}69 & -- & \color{blue}113 &  \color{blue}86 & 	 \color{blue}75			\\
  Rachel & \color{magenta}66 & \textit{\color{magenta}48} & \color{magenta}82 & --	& \textit{\color{blue}67} & \color{blue}88			\\
  Joey & \color{magenta}66 & \color{magenta}85 & \color{magenta}72 & \color{magenta}74 & --	& \color{blue}80			\\
  Phoebe & \color{magenta}78 & \textit{\color{magenta}59} & \color{magenta}60 & \color{magenta}74 & \color{magenta}65 & --				\\
  
 \end{tabular}
\end{table}

As for the other two aggregated graphs, Table~\ref{tab:interactions} shows the number of interactions (normalized by the number of seasons) for s1--s4 (blue) and s5--s10 (magenta) for each pair of friends.
Bold figures indicate high values whereas italics indicate low ones.

We can see that Chandler used to interact a lot with Joey in s1--s4, but this has changed after season 5, when he and Monica started a relationship.
In any case Chandler seems to concentrate his interactions mostly with these two, despite Ross being not only his former college roommate but also, later his brother in law.
As for Monica, in s1--s4 she used to interact more or less the same amount with every other friend (less with Joey).
After season 5 she too really interacts much more with Chandler than with all others.

\section*{\fri\ and other human social networks}

In order to check to what extent \fri\ is similar to other human social networks, a brief comparison was made using data reported in the literature.
Some characteristics of these networks  are  power-law scaling in degree distribution, large clustering coefficients, and a small
mean degrees of separation.

For a comparison, the AE graph of \fri\ is used.
As seen before, it stands out that, although the six friends haver roughly the same degree, there is a big difference to other characters' degrees.
Quantitatively, in terms of frequencies of degree intervals, 0.8\% (the six friends) have degree higher than 1000, another 0.8\% have degree between 100 and 1000 (actually between 100 and 200), 13\% have degree between 10 and 100, and the remaining 85\% have degree smaller than 10.
This way, the degree frequencies is not fully in line with a typical power-law scaling in a human social network, or at least not with one having typical exponent between 2 and 3.

Regarding clustering coefficient, while the AE graph in \fri\ has a value of 0.03 (see Table~\ref{tab:situations}), the following values are reported for other online social networks:   0.16 for the Korean online network Cyworld and 0.26 for MySpace \citep{Ahn+2007};  0.12 for the Japanese Mixi network  \citep{Yuta+2007}; 0.18 for the Dutch Hyves network  and  0.16 for Facebook   \citep{Corten2012}.
It must be noted though that some studies on portions of the Facebook network report a lower value for clustering: 0.0359 as in \url{http://konect.uni-koblenz.de/networks/ego-facebook} (accessed Aug. 26, 2018).
Thus, one can see that \fri' network has a lower clustering coefficient than some online social networks, which is only as expected since it was already mentioned that many characters never get to meet many others. 

Finally, regarding degrees of separation, in \fri\ it is around 3 since virtually all characters are connected to at least one of the six friends.
This value is in line with most of the online social networks: between 3 and 5 \citep{Ahn+2007,Yuta+2007}.
Diameter however, standing at 5, is lower than in other cases. Again, this is a consequence of the fact that many characters are friends with the six friends.

In conclusion, \fri\  cannot be considered a typical human social network, at least not when compared to online social networks using the aforementioned metrics.

\section*{Network structure and communities in \fri}

In  network science, \cd\ corresponds to the task of partition a network into connected subgraphs\footnote{Normally interchangeably referred as groups, communities, and clusters; since the term cluster has a more general meaning in computer science, it is avoided here.}, so that nodes in each subgraph have denser connections (edges) within these subgraphs, as compared to connections with nodes outside them.

Community detection is helpful when analyzing the structure of a network, but it is a challenging problem.
Many  approaches for detecting communities in networks have been designed \citep{Fortunato2010}.
These approaches differ greatly in their underlying philosophy (see \citet{Yang+2016} for a short explanation---as well as their computational complexity---on popular methods that are implemented in the \textit{igraph} package),
Thus, comparison of these different \cd\ methods is not trivial, especially when one does not know the real partition (the ground truth), which is the case in the present study and also in many others.
Besides, it is probably the case that different methods are more or less appropriate for different classes of networks.
In order to help selecting an appropriate method for a given circumstance based on observable properties of a given network, \citet{Yang+2016} provide recommendations for the use of some \cd\ methods.

In order to compare the partition output by each method, some measures have been suggested in the literature, mostly based on similarity measures for clustering and/or on information theory.
Basically, the comparison is pairwise.
In the present work, primarily,  a standard one is used, namely normalized mutual information (\nmi) \citep{Danon+2005}.
\nmi\ is based on the idea that comparing two partitions can be seen as the problem of message encoding: if two partitions are similar, then one needs very little information to infer one from the other.
Thus \nmi\ defines similarity of two partitions as their mutual information normalized by their entropies.
Advantages of \nmi\ are: (i) it does not require that the number of communities in both partitions are the same, (ii) it is bounded in the interval between zero and one.

As pointed out in the literature, not only similarity measures can be used to evaluate the quality of a partition.
Depending on the research questions, if  metadata regarding the particular network is available, this could be used as well to facilitate the interpretation and meaning of the communities found.
Henceforth this will be referred as plausibility assessment.
Finally, a further assessment is made by the so-called mixing parameter $\mu$ that was defined in association with the LFR benchmark \citep{Lancichinetti+2008}.
Be $k_i^{ext}$ the  external degree
of node $i$, i.e., the number of edges connecting $i$ to nodes that belong to different communities, and $k_i^{tot}$ the total degree of node $i$ (as defined before).
Then:
\[ \mu = \frac{k_i^{ext}}{k_i^{tot}} \]

In the LFR benchmark, it is assumed that $\mu$ is  the same for every vertex.
Thus, with abuse of notation, henceforth $\bar{\mu}$ is used to denote the mean mixing parameter of vertices in a graph, as also in \cite{Yang+2016}. 

As a remark, $\mu$ is similar to another measure called embeddedness, 
which is the ratio of the internal degree of a community by its total degree.
The former is given by the sum of the internal degrees of the community's nodes, i.e., twice the number of links inside the community.
The fact that $\mu$ is used here is due to facilitate following the recommendations given in \citet{Yang+2016}.
Notice further that while one expects to see high embeddedness, low $\mu$ is a desirable property.

To investigate the effects of various \cd\ methods  in the \fri\ dataset, three sizes of graphs are shown ahead: 
the graph for a single episode (s4e24), the graph for one whole season (s7), and the AE graph which includes all 236 episodes.
These were selected because they have different sizes and were already introduced and used for the discussion on centralities.

As mentioned, the ground truth about partition of the characters in the show is not known, although metadata allows us to make some plausibility investigations.
In principle, perhaps one would expect that at least some of the \cd\ methods would put the six friends isolated in a single group, with the rest of the characters in other groups, which would be in line with the assumption of surrogate family.
This however does not happen since the six have not only strong connections among themselves, as discussed before, but also  connections with various levels of strength  to many other characters.
Thus, as discussed ahead, no method has produced such a partition.

In the absence of the ground truth, one needs to rely in pairwise comparisons between methods using the \nmi, and also consider plausibility and the value of the mixing parameter $\bar{\mu}$. 
In general, as discussed ahead, if one considers the value of $\bar{\mu}$ as a metric, the best method is label propagation because it yields the lowest $\bar{\mu}$.
However, in general this method produces few communities (the reason behind a low $\bar{\mu}$) that are not totally plausible.
Other methods whose values of $\bar{\mu}$ are low are multilevel, spinglass and leading eigenvector.

The next subsections discuss details about methods and findings for the three sizes of graphs just mentioned.

\subsection*{Community detection in the graph of one episode}

\begin{figure}[t]
 \centering
\includegraphics[width=0.75\linewidth]{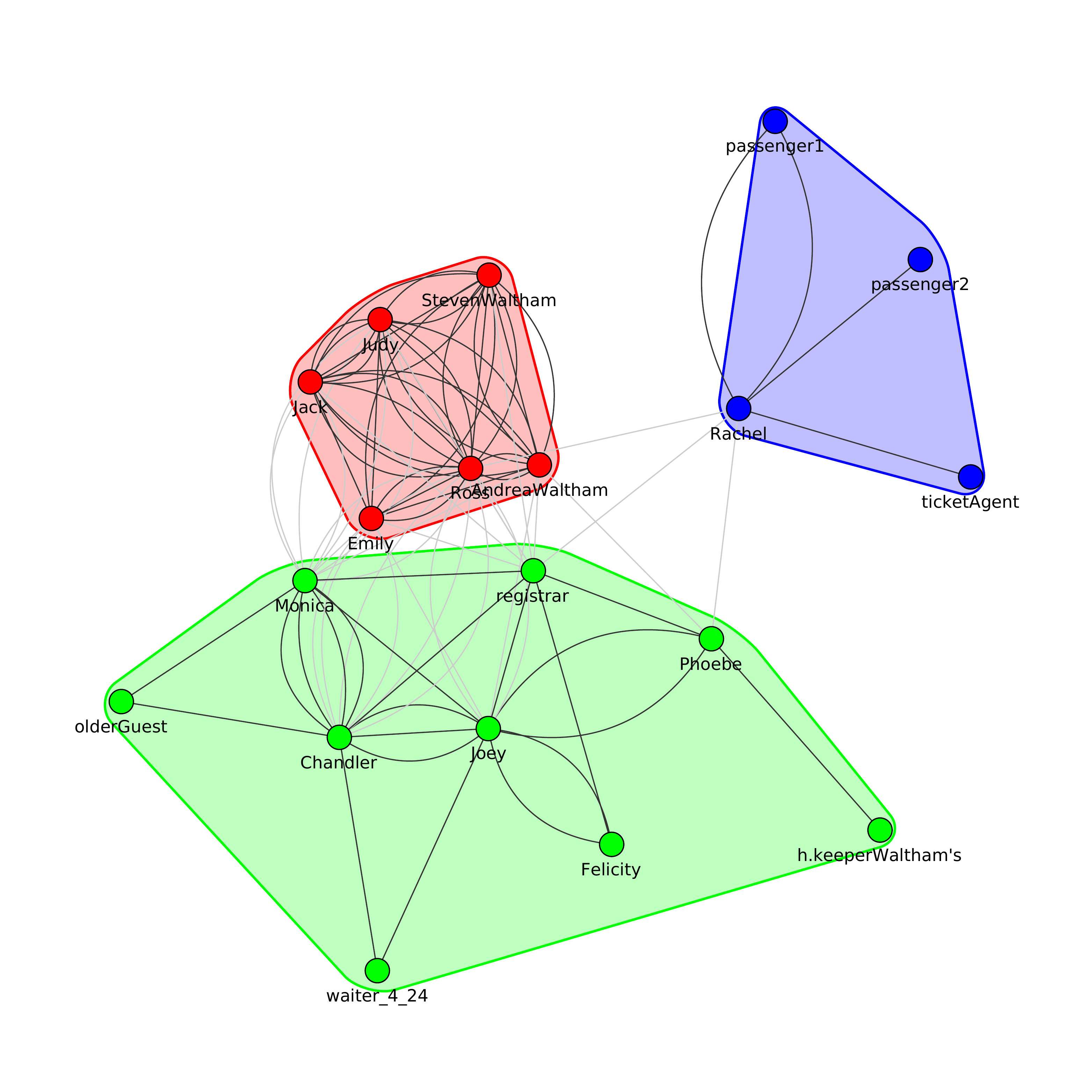}
 \caption{Communities detected in s4e24 using the multilevel method.}
 \label{fig:multilevel}
\end{figure}

\begin{figure}[t]
 \centering
\includegraphics[width=0.75\linewidth]{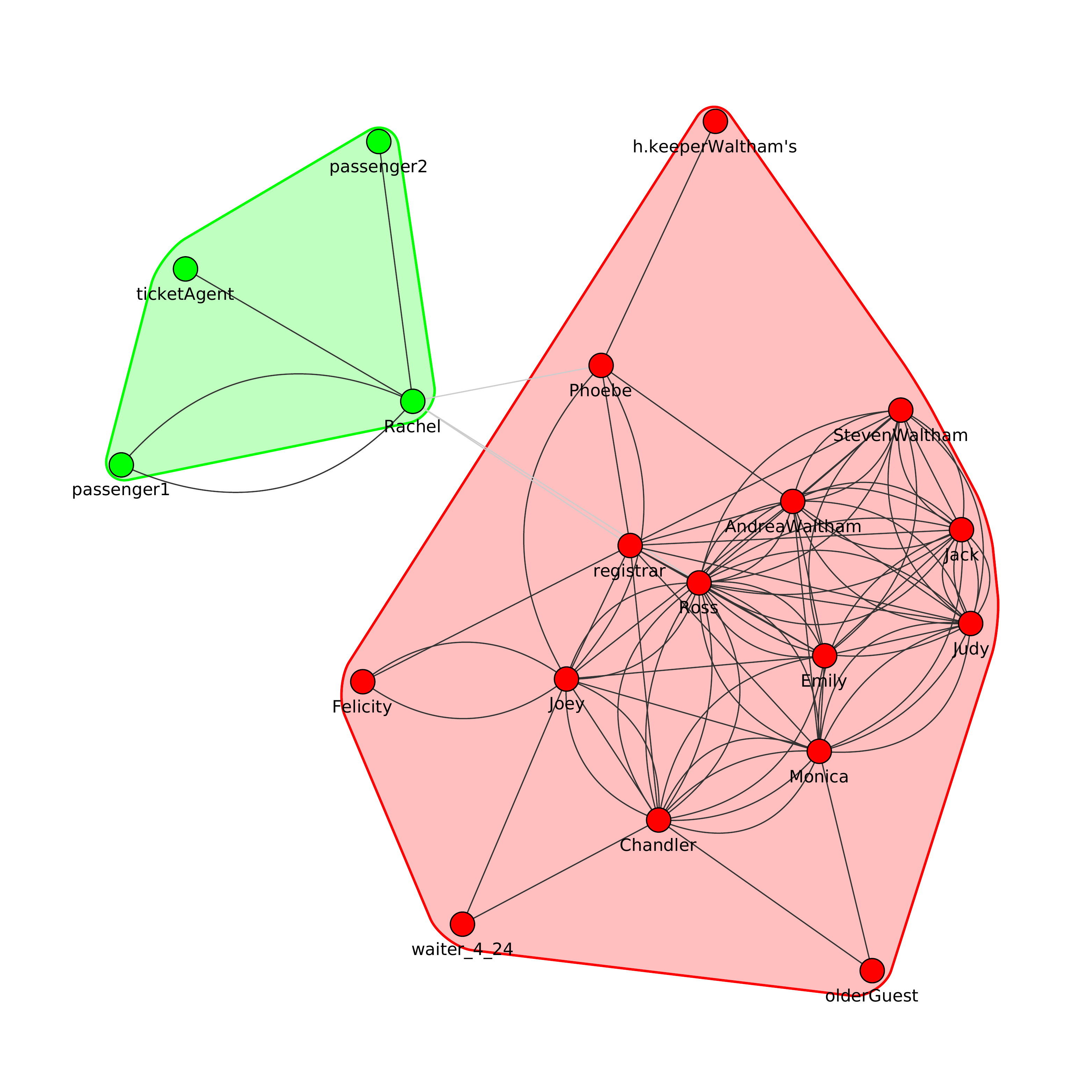}
 \caption{Communities detected in s4e24 using the label propagation method.}
 \label{fig:labelprop}
\end{figure}

\begin{figure}[t]
 \centering
\includegraphics[width=0.75\linewidth]{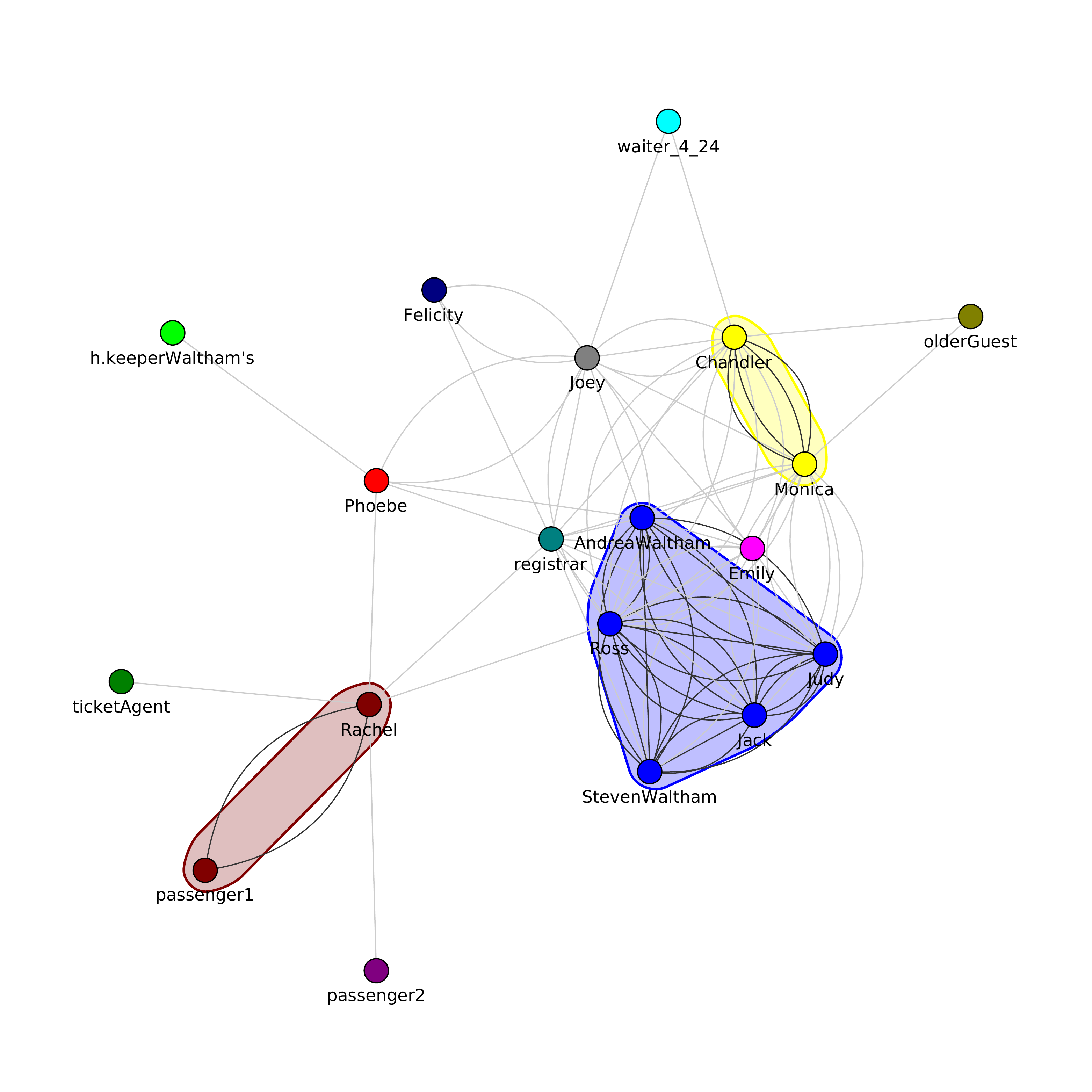}
 \caption{Communities detected in s4e24 using the edge betweeeness method}
 \label{fig:edgebetweenness}
\end{figure}

As said, the recommendations in \citet{Yang+2016} were followed. This  calls for running either multilevel or spinglass to get a first sense of the values of $\bar{\mu}$.
Both produced similar partitions, though multilevel is less expensive and more stable (output does not change, as opposed to spinglass, where stochasticity plays a major role).
In both, the mixing parameter $\bar{\mu}$ is around 0.28, with partitions between three and four communities.
Recall that the lower $\bar{\mu}$, the better.
Fig.~\ref{fig:multilevel} shows the resulting partition into three communities for the multilevel method.
 
\citet{Yang+2016} then suggests running other methods some times to check whether the value of  $\bar{\mu}$ and/or the communities detected are stable.
Given the small size of the graph, all methods were tried.
As general remarks, leading eigenvector performs very similarly to multilevel and spinglass; walktrap and infomap came next (in terms of $\bar{\mu}$), producing more communities (6 to 7); edge betweeeness produced too many communities (13 as in Fig.~\ref{fig:edgebetweenness}, which is high for a single episode),  with a high value of $\bar{\mu} \approx 0.7$; 
finally, as said, label propagation yields a low value of $\bar{\mu}$ (in this case 0.04) but creates just 2 communities,  putting almost all characters in a single one, even if it got characters around Rachel correctly  (see Fig.~\ref{fig:labelprop}).
Several methods got a community composed only or nearly only by Monica and Chandler, which is very plausible for that episode.

\subsection*{Community detection in the graph of one season}

For \cd\ in a bigger graph, the whole season 7 was taken, whose graph was shown in Fig.~\ref{fig:s7}.
As the visualization of the communities found is not always  meaningful---as one can hardly distinguish the partitions---mostly the discussion is textual as follows.

Again, following \citet{Yang+2016}, 
multilevel produced $\bar{\mu}=0.11$, with 4 communities: one  has only three characters in the hospital featured in the show Joey acts on, which makes a lot of sense.
It is worth noticing that this community was found by most of the \cd\ methods applied.
The second small community found by multilevel has two characters in a library.
Then there are two communities that have 25 and 51 characters respectively.
The former includes Monica, Chandler, their parents among others.
The latter includes the other four friends and their acquaintances.
In short, this method produced a very plausible partition for the 81 characters in season 7.

As for other methods, again, label propagation has yield the lowest value for $\bar{\mu}=0.004$ as it produced just two communities. One has those three characters in the hospital but all other characters of the season were put in a single, giant community.
This however makes little sense since characters with completely different centrality---no matter if measured by degree or another kind of centrality measure---are all put together in the same group.
Obviously a very low value of $\bar{\mu}$ does not necessarily indicate a plausible partition.

\begin{figure}[ht]
 \centering
\includegraphics[width=0.9\linewidth]{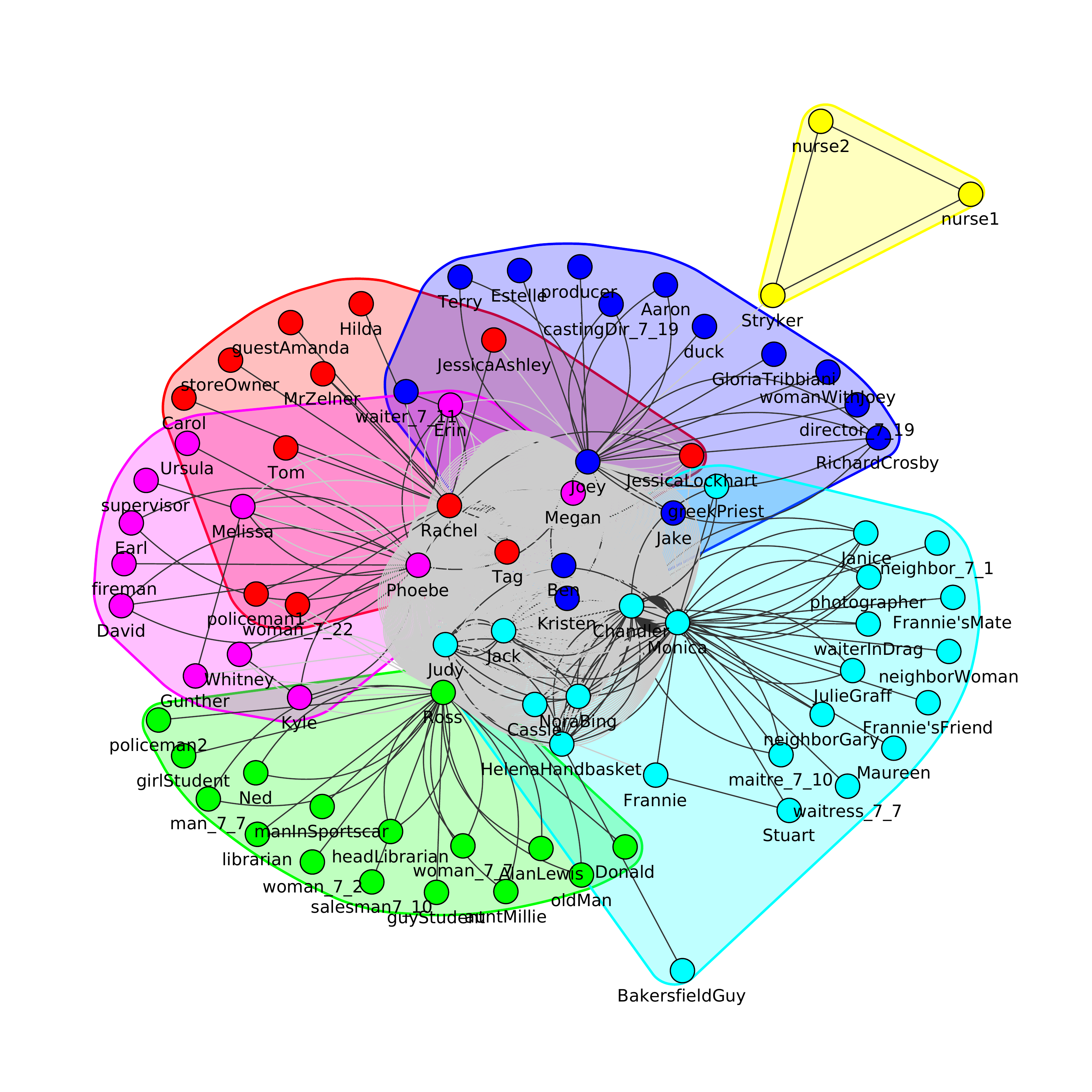} 
 \caption{Communities detected in s7 using the  spinglass method.}
 \label{fig:sgs7}
\end{figure}

One interesting result that is observed 
when applying spinglass to detect communities is that it tends to separate each of  the six friends in its own group.
This can be seen in almost all relatively big graph (e.g., whole season, AE graph).
In season 7 in particular (see Fig.~\ref{fig:sgs7}), 6 groups are created, with $\bar{\mu}=0.21$.
One has the three aforementioned characters in the hospital. Then there is one group for each friend, except that, not surprisingly (as the season revolves around them planning the wedding), Monica and Chandler are in the same group.

Walktrap did not perform well ($\bar{\mu} \approx 0.7$), produced 57 groups. The biggest one includes all 6 friends.
Edge betweeeness, as before, produced a very high number of groups. Basically there is one group with the six friends and Tag, while each other character is  a singleton.

\subsection*{Community detection in the  graph containing all episodes}

Obviously it is not possible to visualize the partition of such a graph as one cannot distinguish the name of characters.
Table~\ref{tab:cdobs} provides 
an overview of all \cd\ methods in terms of $\bar{\mu}$, number of communities, and some observations regarding plausibility.

One can see that methods perform very differently, not only quantitatively ($\bar{\mu}$ and number of groups), but also regarding, e.g., how they partition the six friends: some put them in the same community (which is more or less expected), while some methods have each as ``head of the group", so to say. 

Multilevel remains the best method as it is less expensive, provides a low value of $\bar{\mu}$, and produces plausible communities.
Spinglass is not only much more expensive but also puts each friend in a different community and also creates different communities where recurrent characters like Janice and Gunther are heads of them.
This may be a result of the spin polarity the method is based on.
Leading eigenvector performs more or less as multilevel, but produces less communities.
Again, label propagation creates very few communities (4, which is low for the AE graph).
The other methods all yield high value of $\bar{\mu}$ and at times not plausible communities.
Edge betweeeness is very expensive, produced roughly the same output as walktrap and, as in the previous cases, does not create an acceptable partition, since it has a lot of communities (mostly formed by singletons).

\begin{table}
 \centering
 \caption{All episodes: Overview of performance of \cd\ methods.}
 \label{tab:cdobs}
 \begin{tabular}{p{1.2cm}p{1cm}p{1.7cm}p{6.5cm}}
  \hline
  Method & $\bar{\mu}$ & Communities & Obs. \\ \hline
  ML & 0.17 & 49 & cheap; \textbf{all 6 friends in a relatively big community}; 3 communities with other recurrent characters and family members; rest are in 2-3 per community  \\
  SG & 0.25 & 10 &  expensive; \textit{each friend (plus Janice, Gunther, Pete) in a different community}\\
  LE & 0.34 & 25 & \textbf{all 6 friends in same community}; several plausible communities; second biggest has many non-recurrent characters \\
  LP & 0.003 & 4 & all 6 friends in same community, together with almost all others \\
  IM & 0.50 & 75 & \textit{each friend in a community}; Ross with only other 2 characters; overall not much plausible \\
  WT & 0.79 & 566 & cheap; \textbf{all 6 friends in same community}, together with recurrent characters; rest singletons \\
  EB & 0.92 & 687 & very expensive; produces roughly same as WT \\
  \hline
 \end{tabular}
\end{table}

Finally, all methods were also pairwise compared using \nmi.
Results appear in Table~\ref{tab:nmiAE}.
As a remark, the trend is more or less the same as the comparison of methods regarding season 7, though the values of \nmi\ tend to be lower the bigger the size of the graph because there is always minor disagreements and thus, the bigger the graph, the more disagreements.

What stands up is that there is only one case in which the value of \nmi\ is high, i.e.,  two methods really agree in terms of partitions: walktrap and edge betweeeness.
The others  have very different number of communities (see Table~\ref{tab:cdobs}) and this impacts the values of  \nmi.
Further, label propagation really disagrees with all others since it produces a very low number of communities.
As a remark, another pairwise comparison was made using the adjusted Rand index, with similar results.

\begin{table}[t]
 \centering
 \caption{All episodes: \nmi\ applied to each pair of methods (SG=Spinglass, IM=Infomap, ML= Multilevel, EB=Edge Betweenness, LE=Leading Eigenvector LP=Label Propagation, WT=Walktrap).}
 \label{tab:nmiAE}
 \begin{tabular}{c|cccccc}
  & IM & ML & EB & LE & LP & WT \\ \hline
  SG &  0.43 & 0.35 & 0.46 &  0.33 & 0.023 & 0.43 \\ 
  IM &  & 0.53 & 0.59 & 0.38 & 0.051 & 0.58	\\ 
  ML && & 0.38 & 0.45 & 0.098 & 0.40  \\
  EB  &&&   & 0.33 & 0.026 & 0.92			\\
  LE  &&&& 	& 0.022 & 0.34	\\
  LP   & &&&& 	& 0.030	\\
 \end{tabular}
\end{table}

\section*{Concluding remarks}

In his paper, some well-known facts regarding the sitcom \fri\ were analyzed using techniques from \netth.
Although graphs, even if they are weighted by the number of interactions as here, do not fully account for interactions' intensities since they do not consider length of scene or even emotionally charged interactions, they can be a reasonable indicator.

By doing the math related to graphs obtained by watching the show's episodes, it is possible to come to the following findings.

There is little difference between degrees of the six friends in the same graph; different situations (thus graphs) show different magnitudes of degree centrality; for instance they are  low in season 8 and very high in graphs where basically the six friends interact only among themselves.

Betweenness centrality on the other hand differs a lot for each character, even when aggregated graphs, i.e., a large temporal slices, are considered.

Monica's degree is normally among the highest but her betweenness is  the lowest in most  seasons.
This supports the thesis of her being the queen in her apartment.
Her degree is especially high in Thanksgiving episodes, as expected.
Phoebe's degree and betweenness is normally in the lowest tier.
Although Joey's degree is not very high, his betweenness stands out both overall as well as in many situations.
Ross has the second highest centrality value both in terms of degree as well betweeeness, and he stands out in several situations or time slices.
Rachel, whose role is associated with high popularity,  has neither high degree nor high betweeeness, although she does dominate the last episodes in many seasons. 

While Phoebe's centrality is normally among the lowest of the six friends, it cannot be concluded that her character was ``a little more secondary"  (as initially planned; see \cite{Kauffmann2004}).
And this is definitely not the case for Chandler, whose centrality values are among the highest.

It is also noticeable that many characters never  get to meet many others.
There is a high variance in degrees of the six friends and the rest of the characters, which points to a centralized network \citep{Valente&Fujimoto2010}.

There are strong indications that 
 \fri\ is about a closed group of friends, and, since  birth relationships were not intense, that the six friends are part of  a surrogate family.
 
 
As for the thesis that a trademark of \fri\ is that of stories having (nearly) equal weight within an episode,   it is hard to say that there  is a balance in story threads in single episodes, when degree centrality of the six main characters are considered in a sample of episodes.

Regarding the application of \cd\ methods, an extensive study using different sizes of graphs has shown that different methods yield very different partitions of the characters into groups.
Although the plot of the show can be seen as metadata about these graphs, the real ground truth is not know. Thus the partitions can be evaluated using plausibility coupled with quantitative measures  such as number of groups and the ratio between external and total number of connections of the groups.

This analysis has shown that multilevel is a good method as it is not expensive, has a low ratio between external and total number of connections (thus a desirable property), and has provided plausible partitions in the cases that were evaluated.
In a pairwise comparison, it was also possible to see that there is very little agreement between the partitions produced by the methods.
 
Regarding the temporal aspect, it was shown that graphs for different time slices of the show change.
It remains to be investigated which is the role of the
techniques used by \cite{Prado+2016}, where  specific centrality measures for temporal networks were used.

Also, since gender issues related to \fri\ continue to be the focus of studies (see \cite{Marshall2007,Melcher2017}), one could investigate whether or not there were a gender balance regarding centrality and number of interactions of female and male characters.

\section*{Acknowledgements}

Ana Bazzan is partially supported by CNPq.

\vspace{0.2cm}
\bibliographystyle{chicago}

\end{document}